%% file: main.tex
\documentclass[aps,pre,reprint,groupedaddress]{revtex4-2}
\usepackage{graphicx}
\usepackage{amsmath}
\usepackage{dcolumn}
\usepackage{amssymb}
\usepackage{color}
\usepackage{lineno}
\usepackage{hyperref}
\usepackage{float}
\usepackage{xcolor}
\usepackage{url}
\hypersetup{
    colorlinks,
    linkcolor={red!50!black},
    citecolor={blue!50!black},
    urlcolor={blue!80!black}
  }

\def\SH{SH\ }

\input{./hudef}\def\pdep{{\tt pde2path}}\def\auto{{\tt AUTO}}
\def\mlab{{\tt MATLAB}}
\begin{document}
\preprint{APS/123-QED}

\title{Localized and extended patterns in the cubic-quintic Swift-Hohenberg equation \\ on a disk}
\author{Nicol\'as Verschueren}
\email{nverschueren@berkeley.edu}
\affiliation{Physics Department, University of California at Berkeley, Berkeley CA 94720, USA}
\author{Edgar Knobloch}
\email{knobloch@berkeley.edu}
\affiliation{Physics Department, University of California at Berkeley, Berkeley CA 94720, USA}
\author{Hannes Uecker}
\email{hannes.uecker@uni-oldenburg.de}
\affiliation{Institute for Mathematics, Carl von Ossietzky University of
  Oldenburg, Oldenburg, Germany}

\date{\today}

\begin{abstract}
Axisymmetric and nonaxisymmetric patterns in the cubic-quintic Swift-Hohenberg
equation posed on a disk with Neumann boundary conditions are studied via
numerical continuation and bifurcation analysis.  Axisymmetric localized
  solutions in the form of spots and rings known from earlier studies
  persist and snake in the usual fashion until they begin to interact
  with the boundary. Depending on parameters, including the disk radius,
  these states may or may not connect to the branch of domain-filling target states.
Secondary
  instabilities of localized axisymmetric states may create multi-arm
  localized structures that grow and interact with the boundary before
  broadening into domain filling states. High azimuthal wavenumber
  wall states referred to as daisy states are also found. Secondary bifurcations from these
  states include localized daisies, i.e., wall states 
  localized in both radius and angle. 
Depending on
  parameters, these states may snake much as in the one-dimensional
  Swift-Hohenberg equation, or invade the interior of the domain, 
yielding states referred to as worms, or domain-filling stripes. 
\end{abstract}


\maketitle
 \input{body}

\begin{acknowledgments}
The work of NV and EK was supported in part by
the National Science Foundation under grant DMS-1908891. 
NV was also funded by the National Agency for Research and Development
(ANID) through the Scholarship Program: Becas de Postdoctorado en el
extranjero, becas Chile 2018 No. 74190030.
\end{acknowledgments}

\bibliography{bibi}

\end{document}

%% file: hudef.tex
\def\bmip{\begin{minipage}{\textwidth}}\def\emip{\end{minipage}}
\def\huga#1{\begin{gather} #1 \end{gather}}

\def\ig{\includegraphics}\def\sm{\small}
\newcommand{\btab}[2]{\begin{tabular}{#1}#2\end{tabular}}
\def\rb{\raisebox}

\newcommand{\bsmm}{\begin{large}\left(\begin{smallmatrix}}
\newcommand{\esmm}{\end{smallmatrix}\right)\end{large}}

\def\pa{{\partial}}
\newcommand{\bce}{\begin{center}}\newcommand{\ece}{\end{center}}
\newcommand{\bci}{\begin{compactitem}}\newcommand{\eci}{\end{compactitem}}
\newcommand{\bcen}{\begin{compactenum}}\newcommand{\ecen}{\end{compactenum}}
\newcommand{\reff}[1]{(\ref{#1})}

\newcommand{\hs}[1]{{\hspace{#1}}}\newcommand{\vs}[1]{{\vspace{#1}}}

\def\eps{\varepsilon}

\newcommand{\barr}{\begin{array}}\newcommand{\earr}{\end{array}}
\newcommand{\bpm}{\begin{pmatrix}}\newcommand{\epm}{\end{pmatrix}}
\newcommand{\bsm}{\left(\begin{smallmatrix}}
\newcommand{\esm}{\end{smallmatrix}\right)}
\newcommand{\ba}{\begin{array}}\newcommand{\ea}{\end{array}}
\def\dd{\, {\rm d}}

\def\Om{\Omega}

\def\bd{\begin{displaymath}} \def\ed{\end{displaymath}}
\def\ba{\begin{array}} \def\ea{\end{array}}
  
\def\eps{\varepsilon}


%% file: body.tex
\section{Introduction}

Pattern formation is a familiar feature of many physical, chemical and
biological systems. Patterns generally form as a result of a
symmetry-breaking instability of a spatially homogeneous state. The
simplest situation arises on an infinite domain in one, 
two, or three space dimensions (1D, 2D or 3D, respectively), 
since this formulation admits spatially
periodic structures. These persist when the domain is replaced by
a periodic domain, or with some restrictions, by a domain with Neumann
boundary conditions. The subject is reviewed in the books by Hoyle
\cite{hoyle} and Cross and Greenside \cite{greenside}. However, in
most applications, particularly in fluid dynamics and chemical
systems, the presence of lateral boundaries becomes of fundamental
importance. This paper is devoted to the explanation of pattern
formation on perhaps the simplest bounded domain in 2D, the finite disk,
focusing on phenomena associated
with subcritical instabilities of the homogeneous state, i.e., associated
with the presence of bistability between the homogeneous state and 
different pattern states. 

For this purpose we select the simplest pattern-forming equation, the 
Swift--Hohenberg (SH) equation, a dissipative evolution equation for
a scalar field $u({\bf x},t)$ in 2D. The equation is characterized by a
finite wave number instability that takes place, in an infinite system,
at $\eps=0$, where $\eps$ is the bifurcation
parameter. Because of its variational structure, all time-dependence
ultimately dies out, allowing us to focus on time-independent
solutions. We choose the nonlinear terms to be of cubic-quintic type
in order to allow for bistability between the homogeneous state $u=0$
and a stripe-like pattern. This type of equation arises as a plausible
model of many systems, particularly those arising in fluid dynamics,
although a rigorous derivation is lacking \cite{ma}. However, because
of its simplicity the equation has been the model of choice for many
pattern formation studies, and this is the case in the present
contribution as well.

\begin{figure*}
  \begin{center}
    \ig[width=.8\textwidth]{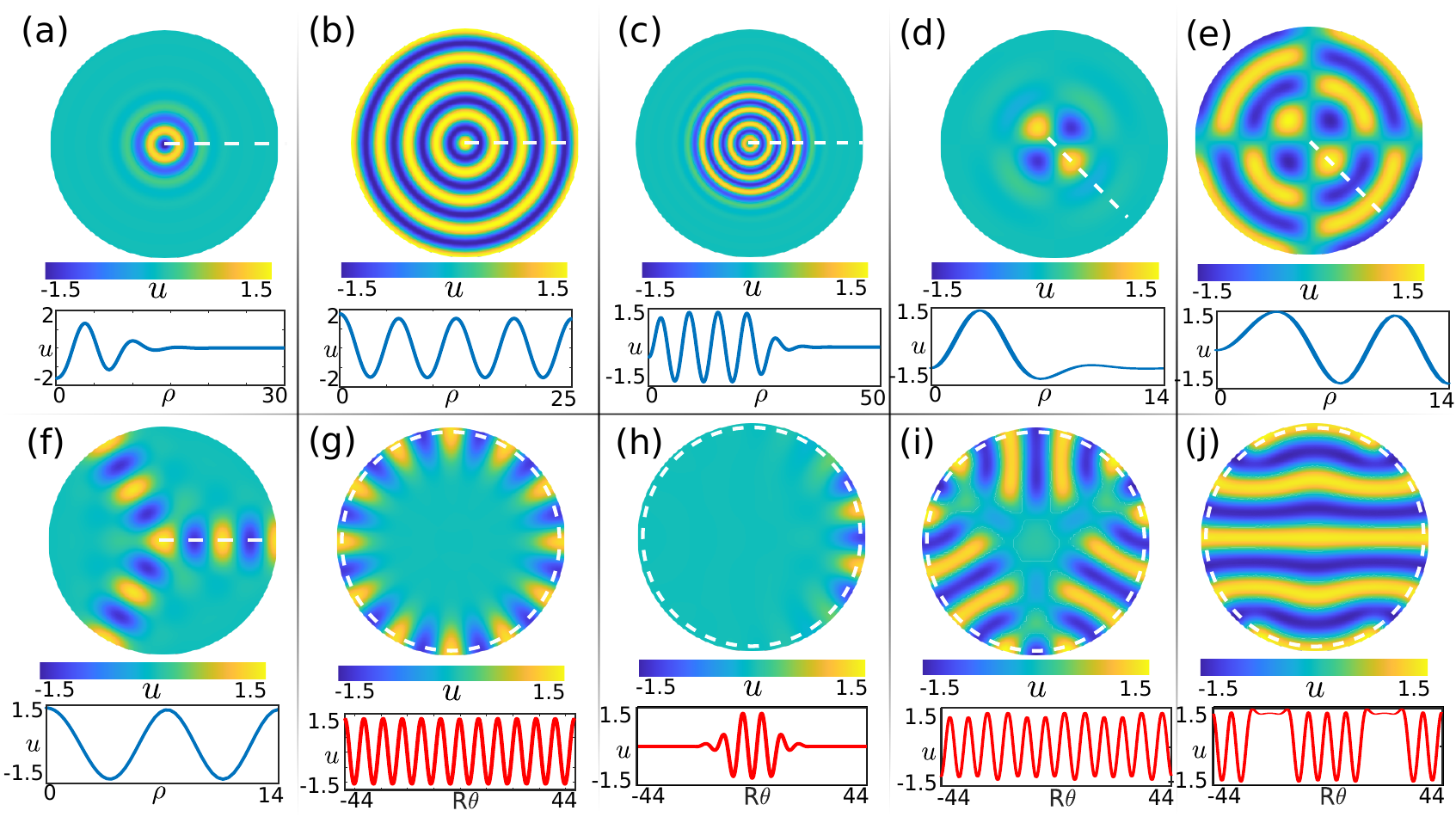}
      \end{center}
  \caption{States observed in Eq. (\ref{eq:sh35}) with
    the boundary conditions (\ref{eq:bc1}), $q=1$.
    (a) Spot $(R,\eps,\nu)=(30,-0.580,2)$,
    (b) target $(R,\eps,\nu)=(25,-0.780,2)$,
    (c) ring state $(R,\eps)=(50,-0.650,2)$,
    (d) $4^{-}$-arm localized state $(R,\eps,\nu)=(14,-0.5980,2)$,
    (e) $4^{-}$-arm extended state $(R,\eps,\nu)=(14,-0.6460,2)$,
    (f) $3^+$-arm state $(R,\eps,\nu)=(14,-0.7048,2)$,
    (g) daisy state $(R,\eps,\nu)=(14,-0.353,1.5)$,
    (h) plucked daisy state $(R,\eps,\nu)=(14,-0.3231,1.5)$,
    (i) worm-like state $(R,\eps,\nu)=(14,-0.2755,1.4)$, and
    (j) stripe-like state $(R,\eps,\nu)=(14,-0.6,2)$.
    Each state is accompanied in a lower panel by a 1D profile along the dashed white line in the panel above; in (g)-(j) this line is slightly displaced from the perimeter for ease of visibility. }
  \label{fig:solutions}
\end{figure*}

In the case of a finite disk, two basic types of localized states occur: states localized in the interior of the
domain, and states localized near its boundary. The former behave much
like analogous structures in the plane, at least until they have grown
to such a size that they begin to interact with the boundary. In
contrast, the latter, referred to here as wall states, are present only because of the
boundary, and so have no analogue in the infinite plane. We
focus here on understanding the basic properties of these solutions
and their interaction as parameters are varied, as well as on their
interaction with coexisting domain-filling structures such as
target states and other stripe-like states. The former are parallel to the
boundary, while the latter are frequently perpendicular to it. 
Periodic wall states are also of special interest. We refer to
these as daisy states, and since these also bifurcate subcritically,
we expect the presence of azimuthally localized daisy states resembling 
partially plucked daisies. We find that these states do indeed exist and that they
snake, for moderate subcriticality, much like the localized states in the
1D cubic-quintic \SH equation. However, this
ceases to be the case for stronger subcriticality, for which the
localized daisies expand into the domain interior instead of 
expanding along the boundary. Figure \ref{fig:solutions} illustrates some of
the solutions studied in this paper. 

Some of our pattterns resemble states
generated in supercritical steady state bifurcations from a trivial
state on a disk, for instance in convection in a cylinder
\cite{crosshohenberg,hof,tuckerman1,tuckerman2,ahlers,ma1} and in
flame dynamics above a circular burner \cite{Palaciosbruss}. However,
in these systems spatial localization is generally absent, 
because roll convection sets in supercritically (the midplane or
Boussinesq symmetry of this system precludes subcritical hexagons at
onset) while combustion away from onset also behaves like a
supercritical system. Near onset, however, ignition is often
subcritical, a fact that may be responsible for the appearance of
localized hotspots and flickering in this regime
\cite{lojacono3}. Related structures are also seen in reaction-diffusion
systems on a disk \cite{sheintuch}, and in vertical cavity surface
emitting lasers (VCSELs) with a round aperture \cite{krauskopf}.

Localized wall modes are present in rotating
convection in a cylinder \cite{goknme93,zhong}; these modes precess
in a retrograde fashion, and in the strongly nonlinear regime appear
to be responsible for the boundary zonal flows observed in
experiments at high Rayleigh number (strong forcing)
\cite{favier}. Similar rotating states are present even in
nonrotating Rayleigh-B\'enard convection but appear via a
symmetry-breaking Hopf bifurcation from an axisymmetric state
\cite{tuckerman3}. However, such dynamical states cannot be described
by variational systems such as the \SH equation with
Neumann boundary conditions, although both mixed (i.e., Robin)
boundary conditions \cite{GolKnospirtar} or the so-called spiral
boundary conditions \cite{dellnitz} admit the presence of Hopf
bifurcations and hence persistent dynamics.
 
The remainder of the paper is organized as follows. Section
\ref{sec:model} introduces the model, emphasizing 
symmetries, effective number of parameters, and variational structure. 
In addition, linear
stability properties of the $u=0$ state are determined and thresholds
for different types of instabilities are identified. Sections
\ref{sec:axi} and \ref{sec:nonaxi} describe the results of numerical
continuation of some of the steady states generated by these unstable
modes, focusing on two distinct types of solutions, namely
axisymmetric solutions [Fig.~\ref{fig:solutions}(a)-(c)] and nonaxisymmetric solutions that include $m$-arm states
[Fig.~\ref{fig:solutions}(d)-(f)], wall-mode states [Fig.~\ref{fig:solutions}(g),(h)] and worms and stripes
[Fig.~\ref{fig:solutions}(i),(j)].
The axisymmetric states are studied in Sec.~\ref{sec:axi} via solutions of a nonautonomous boundary value problem
in the radial coordinate $\rho$, and their relation to the results on
an unbounded domain \cite{radlloyd,mccalla,bramburger} is explored.
Section \ref{sec:nonaxi} describes the
organization of nonaxisymmetric solutions on the disk, which we 
classify into multiarm solutions and daisy
states. Bifurcations from daisy states yield localized daisies, 
and for moderate subcriticality the associated branches closely resemble 
the classical 1D snakes-and-ladders scenario \cite{burkesh351}. However, these snakes break up 
with increased subcriticality, leading to worm states at the wall. 
Section \ref{sec:energy} shows a comparison between the solution branches 
in terms of the free energy, while Sec.~\ref{sec:otherpatterns} illustrates additional patterns that are also
present. The paper concludes in Sec.~\ref{sec:discussion}
with a discussion and suggestions for future work. 
An Appendix contains some details on the numerics, and 
further information, including movies stepping through some of the 
bifurcation diagrams, is provided in the Supplementary Information (SI).

\section{The model}
\label{sec:model}

\begin{figure}[H]

   \begin{center}
      \ig[width=8.6cm]{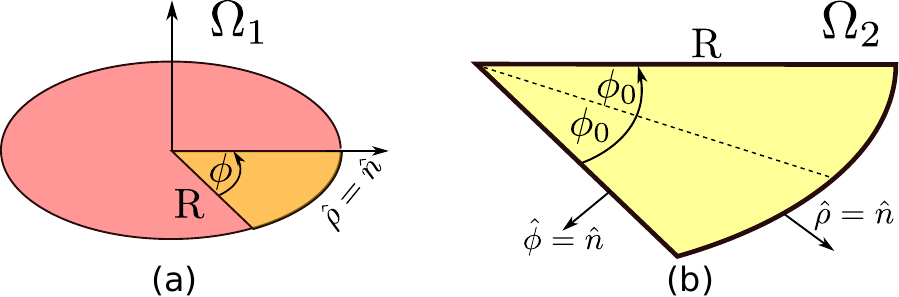}
   \end{center}
   \caption{The domains $\Omega_1$ and $\Omega_2$ defined in Eqs.~(\ref{eq:dom}) and (\ref{eq:dom1}) showing the normal vector $\hat {\bf n}$.
        }
      \label{fig:discos}
   \end{figure}

We consider the \SH model with a cubic-quintic nonlinearity,
\begin{equation}
  \label{eq:sh35}
  \partial_t u=\eps u+\nu u^3-u^5-(q^2+\Delta)^2 u,
\end{equation}
where $u(\mathbf{x},t)$ is a real-valued scalar field and
$\eps$, $\nu$ and $q$ are parameters. In contrast to much of the earlier
literature, here the equation is posed on a disk of radius $R$ subject to the boundary conditions
\begin{equation}
   \label{eq:bc1}
   \left. \nabla u \cdot \hat n\right|_{\delta \Omega}=0,\quad \left. \nabla (\Delta u)\cdot \hat n\right|_{\delta \Omega}=0, 
 \end{equation}
referred to as Neumann boundary conditions. We solve this problem on two distinct domains, a finite disk $\Omega_1$ of radius $R$ illustrated in Fig.~\ref{fig:discos}(a) and a sector $\Omega_2$ of opening angle $2\phi_0$ illustrated in Fig.~\ref{fig:discos}(b):
 \begin{subequations}
  \begin{align}
   \label{eq:dom}
   \Omega_1&=\{\mathbf{x}\in (\rho,\phi)\in \mathbb{R}^2| \rho\in[0,R],
             \phi\in[0,2\pi]\},\\
   \label{eq:dom1}
  \Omega_2&=\{\mathbf{x}\in (\rho,\phi)\in \mathbb{R}^2| \rho\in[0,R], \phi\in[-\phi_0,\phi_0]\}.
  \end{align}
   \end{subequations}
   
In the following the symmetry $u\to-u$ of Eq.~(\ref{eq:sh35}) will play a fundamental role. Although the problem is specified by four parameters, $\eps,\nu,q,R$, we can without loss of generality set $q=1$ since other values of $q$ can be accommodated by the rescaling 
\begin{equation}
  \label{eq:scali}
  \Delta=q^2\Delta',\; u=qu',\;
  \partial_t=q^4\partial_{\tau},\;
  \eps=\eps'q^4,\; \nu=\nu' q^2.
\end{equation}
Thus $R$ is measured in units of $q^{-1}$.

The model (\ref{eq:sh35}), posed on the finite disk $\Omega_1$ with
the boundary conditions (\ref{eq:bc1}), possesses variational
structure. More precisely, the free energy
 \begin{equation}
   \label{eq:lyapfunc}
       \mathcal{F}[u]\equiv\int_\Omega
    \left(\frac{1}{2}[(q^2{+}\Delta)u]^2{-}\frac{\eps}{2} u^2
{-}\frac{\nu}{4}u^4{+}\frac{1}{6}u^6 \right) d\mathbf{x},
  \end{equation}
satisfies, on using integration by parts and the boundary conditions, 
     \begin{align}
   \label{eq:dfdt}
   \frac{d\mathcal{F}}{dt}
=& \int_{\Omega} \left[-\eps u-\nu u^3+u^5+q^2((q^2+\nabla^2)u) 
\right](\partial_t u)\,  d\mathbf{x} \nonumber \\ &
+ \int_{\Omega} ((q^2+\nabla^2)u)\nabla^2 
(\partial_t u)\, d\mathbf{x}\nonumber\\
=&-\int_{\Omega_1} (\partial_t u)^2 d\mathbf{x}. 
     \end{align}
Since $\mathcal{F}[u]$ is bounded from below, it decreases along trajectories until $\partial_t u=0$.
The variational structure of Eqs.~(\ref{eq:sh35})--(\ref{eq:bc1}) thus rules out persistent dynamics
and we therefore focus on time-independent solutions of the problem. 

Our main tool for exploring time-independent solutions of
  (\ref{eq:sh35}) is numerical continuation. We make
  extensive use of the packages \auto\ \cite{auto} and \pdep\ 
  \cite{pde2path,hannesbook,p2phome}. 
For numerical continuation in 2D, the number of states often increases
rapidly with the size of the domain and branch jumping, i.e., uncontrolled and undetected switching of states from one solution 
branch to another, becomes a problem in the neighborhood of bifurcation points.
One strategy to mitigate this problem is to employ symmetries to restrict the study to a smaller
domain, a procedure that also reduces numerical effort. See 
\cite[\S3.6.1, \S8.3.1]{hannesbook} for further comments.  
Here we fix a moderate value for the disk radius and in some settings study 
the problem on the sectorial domain (\ref{eq:dom1}) instead of the whole disk
(\ref{eq:dom}). For the opening angle $2\phi_0$ we take $\phi_0=\pi/n$ with $n$ an integer
and impose Neumann boundary conditions along $\phi=\pm\phi_0$ in addition to $r=R$.

\subsection{Linear stability of the trivial state $u=0$}\label{sec:lsa}

The linearization of (\ref{eq:sh35}) about the homogeneous solution $u=0$ reads
  \begin{equation}
    \label{eq:sh35n0}
    \partial_tv=[\eps-(q^2+\Delta)^2]v,\quad |v|\ll 1.
  \end{equation}
Equation (\ref{eq:sh35n0}) can be solved via separation of variables, using the eigenfunctions of the Laplacian in polar coordinates $(\rho,\phi)$, i.e., we seek solutions of the form $v(\rho,\phi,t)=e^{\sigma t}w(\rho,\phi)$ with 
\begin{equation}
  \label{eq:sh35nosol}
  w(\rho,\phi)=J_m(k \rho)\cos(m\phi+\alpha),
\end{equation}
where $\sigma$, $\alpha$ and $k$ are constants to be determined for each value of the azimuthal wave number $m\in\mathbb{N}_0$. Substituting $v(\rho,\phi,t)$ into (\ref{eq:sh35n0}) yields the dispersion relation
\begin{equation}
  \label{eq:drk}
  \sigma=\eps-(q^2-k^2)^2.
\end{equation}
Instability sets in when $\sigma$ crosses zero. The resulting critical
case $\sigma=0$ admits four values of $k$,
$k_{\pm,\pm}=\pm\sqrt{q^2\pm\sqrt{\eps}}$. However, with $m \in
\mathbb{N}_0$ it follows that $J_m (-x)=(-1)^m J_m (x)$ and
consequently that only two of these solutions are linearly
independent. Thus
\begin{equation}
  \label{eq:soleven}
w(\rho,\phi)=(A_m J_m (k_{+}\rho)+B_m J_m(k_{-}\rho))\cos(m\phi+\alpha)
\end{equation}
where $k_{\pm}=\sqrt{q^2\pm \sqrt{\eps}}$. In order to determine the constants $A_m$, $B_m$ and $\alpha$, we impose the boundary conditions (\ref{eq:bc1}). In the case of the full disk (domain $\Omega_1$), the solution must be periodic in the angular variable $\phi$, implying that $m \in \mathbb{N}_0$. In the case of a slice of half-angle $\phi_0=\pi/n$ (domain $\Omega_2$ with $1< n\in \mathbb{N}$), the boundary conditions in the angular coordinate yield 
$$\sin\left(m\frac{\pi}{n}+\alpha\right)=0,$$
which can be satisfied by either
\begin{equation*}
  \alpha=0,\; m=ln,\; \text{or} \; \alpha=\frac{\pi}{2},\;  m=\frac{(2l-1)n}{2},\; l\in\mathbb{N}_0,
\end{equation*}
where without loss of generality we take $l=1$. We interpret this result as follows. Either $n=m$ in which case $n$ copies of the solution generate an $m$-fold solution on the whole disk, or $n=2m$ implying that one must first double the domain via reflection in $\phi=\phi_0$ or $\phi=-\phi_0$ before replicating the result to obtain an $m=n/2$-fold state \cite{crawford}. Thus $n$ is an even integer, and solutions of this type with $n$ odd are not solutions on the whole disk.

The boundary conditions at $\rho=R$ are the same for both the full disk and the sector: these yield a set of two linear equations for $A_m$, $B_m$ given by
\begin{equation*}
\mathbb{M} \left(\begin{array}{c} A_m \\B_m \end{array}\right)
{=}\left(\begin{array}{c}0  \\0 \end{array}\right),\;\; 
\mathbb{M}{\equiv} \left(\begin{array}{cc}
        k_+J_m'(k_+R)   &     k_-J_m'(k_-R)\\ 
 k_+^3J_m'(k_+R)   &  k_-^3J_m'(k_-R) \end{array}\right).           
\end{equation*}
To find nontrivial solutions of this equation for given $m$ and $R$ we 
determine $\eps$ from
\begin{equation}
  \label{eq:condi}
  {\rm det}(\mathbb{M})=2\sqrt{\eps}\sqrt{q^4-\eps}J_m'(k_+R)J_m'(k_-R)=0.
\end{equation}      
If $\eps=0$, then $k_-=k_+>0$, and condition (\ref{eq:condi}) is satisfied trivially. In this case
Eq.~(\ref{eq:drk}) shows that no instability takes place. Similarly, $\eps=q^4$ also satisfies (\ref{eq:condi}) for all $m$. We disregard this case because we are interested in small values of $\eps$ and $q\sim O(1)$. Condition (\ref{eq:condi}) thus reduces to
$$J_m'(k_+R)J_m'(k_-R)=0.$$
In order to illustrate the predictions from this condition, we consider
the cases $q=1$, $\phi_0=\frac{\pi}{2}$ (half disk) and $R=14$ and $R=15$ (Fig.~\ref{fig:linpred}).
Using computer algebra we find the first six values of $\eps$ 
satisfying \reff{eq:condi}. On the full disk, we obtain the same 
eigenvalues, but for $m\ne 0$ these are of double multiplicity. 

\input{lpf2_ek}

The values found are in good agreement with the bifurcation points found
from one-parameter numerical continuation of the trivial state $u=0$
of (\ref{eq:sh35}) on the half disk. The percentage difference
between the predicted and observed values of $\eps$ is always
under 1\%. Figure \ref{fig:linpred} illustrates the
first six predicted (and numerically computed) eigenvectors when
$R=14$ and $R=15$, sorted by the corresponding eigenvalues $\eps$.
Remarkably, the only mode that is common to both lists is $m=3$,
indicating a strong dependence on $R$ of the initial bifurcation sequence.
Although the number of possibilities increases rapidly with the domain size,
we can always identify two qualitatively different types of eigenvectors:
\textsl{Axisymmetric} solutions characterized by $m=0$ (when $R=14$ these
appear at $\eps_5\approx8.88\times 10^{-3}$, Fig.~\ref{fig:linpred}(a);
when $R=15$ the onset of axisymmetric states is delayed to $\eps_{13}\approx4.20\times10^{-2}$
and $\eps_{14}\approx4.50\times10^{-2}$, i.e., two such modes arise in close succession) and
\textsl{wall modes} characterized by a high wave number $m$
and confined to the periphery $\rho\sim R$ of the domain [e.g., $\eps_3\approx2.92\times10^{-4}$ at $R=14$ and
$\eps_1\approx8.6\times10^{-5}$ at $R=15$, Figs.~\ref{fig:linpred}(a) and (b)].
The organization, stability and
interconnections among the solution branches spawned by these modes, 
radial and wall, are discussed in the following sections. 
We start with these modes because
(a) the radial mode is essentially 1D and, on the infinite disk and for 
the quadratic--cubic case, has been studied before; 
(b) the daisy mode is expected to be associated with quasi-1D snaking 
of localized daisies. Additionally, both show interesting further 
bifurcations. These two states are thus a natural starting 
point for discussing the organization of the very rich solution structure
that exists even on small to moderate size disks.

\section{The axisymmetric case}\label{sec:axi}
    \subsection{Boundary value formulation}
    \label{sec:contibvp}

The steady state problem for axisymmetric solutions $u=u(\rho)$, $\rho \in[0,R]$, 
with the boundary conditions (\ref{eq:bc1}) can be written as the
nonautonomous 1D boundary value problem 
\begin{subequations}
  \label{eq:bvprad}
  \begin{align}
    v_1' &=  v_3,\\
    v_2' &=  v_4,\\
    v_3' &=v_2-q^2v_1-\frac{v_3}{v_5},\\
    v_4' &=\eps v_1+\nu v_1^3-v_1^5-q^2 v_2-\frac{v_4}{v_5},\\
    v_5' &=1,
    \end{align}
  \end{subequations}
where $v_i'=\frac{d v_i}{d\rho}$, $i=1,\dots,5$. The connection between the components of
$\mathbf{v}\equiv (v_1,v_2,v_3,v_4,v_5)$ and $u$ is as follows:
$$ v_1=u,\quad v_2=(q^2+\Delta)u,\quad v_3=u_1',\quad v_4=u_2', \quad v_5=\rho.$$
The system~(\ref{eq:bvprad}) is not the only choice for
studying steady axisymmetric solutions of Eq.~(\ref{eq:sh35}) as
a boundary value problem, and similar formulations have been used before 
to explore the quadratic-cubic \SH model \cite{radlloyd,mccalla,bramburger}.
The advantage of~\reff{eq:bvprad} is that the 
boundary conditions~(\ref{eq:bc1}) for the system on the disk yield 
the uncoupled boundary conditions 
  \begin{subequations}
    \label{eq:bvpbc}
    \begin{align}
    \left.v_3\right|_{\rho=0,R}&=0=\left.\partial_{\rho}
      u\right|_{\rho=0,R}=\left.\nabla u \cdot \hat
      \rho\right|_{\partial\Omega},\\
    \left.v_4\right|_{\rho=0,R}&=0=\left.\partial_{\rho}(q^2u+\Delta
      u)\right|_{\rho=0,R}=\left.\nabla(\Delta u)\cdot \hat \rho
                              \right|_{\partial \Omega},\\
      \left.v_5\right|_{\rho=0}&=0=\left.\rho\right|_{\rho=0}, 
    \end{align}
  \end{subequations}
which allow a simple numerical implementation, for instance, using the boundary value
routine of \auto, even for large $R$. This approach does
not provide linear stability information, however, unless one solves
the linearized problem in parallel (see e.g., \cite{uwe,jens1,jens2,uwetut}).

Since we are also interested in 
continuation of the branches that bifurcate from axisymmetric states,
we also compute such states in 2D (albeit for moderate $R$) 
using \pdep, which yields linear stability information without additional effort.

\subsection{Axisymmetric solutions on a small disk}

\begin{figure}
  \begin{center}
     \ig[width=8.6cm]{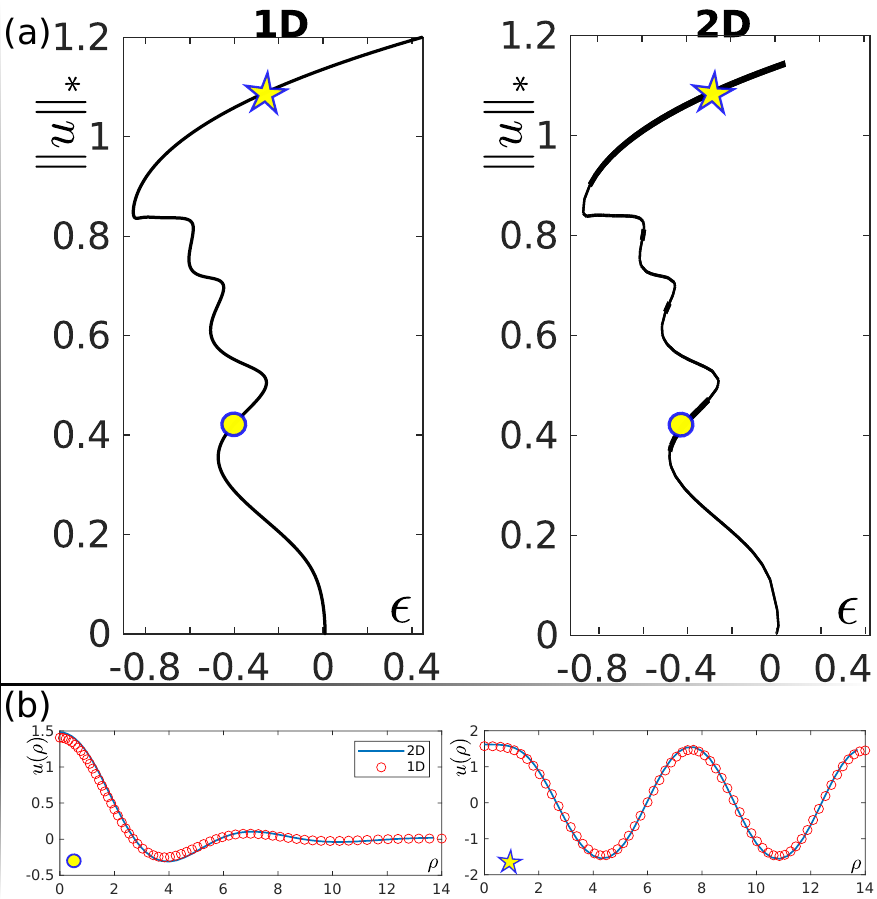}
      \end{center}
      \caption{(a) Comparison between the 1D (\auto, left) and 
2D (\pdep, right) continuation of axisymmetric states on a disk of radius
        $R=14$ when $\nu=2$ and $q=1$. Thick line segments in the 2D case indicate stable solutions.
(b) Radial profiles of representative spot and target
        states at locations indicated in (a) by an open circle and a star.
        Open red circles correspond to 1D computations
        while the solid blue line shows the corresponding profile obtained from 2D computations.}
  \label{fig:compabeta0}
\end{figure}

We begin by demonstrating the equivalence between the \auto\
solution of the 1D boundary value problem (\ref{eq:bvprad}) and
the \texttt{pde2path} solution of the 2D problem (\ref{eq:sh35}). In Fig.~\ref{fig:compabeta0}
we compare the {\it spot} and {\it target} patterns computed by both procedures when
$R=14$, $\nu=2$, $q=1$. Panel (a) shows the branch of axisymmetric
solutions emerging from the trivial branch $u=0$ at $\eps=\eps_5$ in the
1D and 2D formulations. To present the solution branches we employ the quasi-1D norm
\huga{\label{u2def} 
  \|u\|_*=\sqrt{\frac {1}{R}\int_0^R u^2(\rho)\,d\rho}\,.
}
This norm is preferred since it avoids suppressing {\it spots} at the center of the disk,
in contrast to the 2D $L^2$ norm used later for nonaxisymmetric states. Panel (b) shows the
radial profiles of two representative axisymmetric solutions at locations indicated by a circle
and a star in the bifurcation diagrams in (a). These correspond to a spot and a target state, respectively.
Because of the symmetry $u\to -u$ of the cubic-quintic \SH equation, 
there is
only one type of spot in this system, in contrast to the quadratic-cubic \SH equation \cite{mccalla}.

In contrast to the 1D problem on the real line \cite{burkesh351},
in the axisymmetric case described by \reff{eq:bvprad} the $u=0$ branch becomes unstable to an 
already localized state given by a Bessel function of the first kind of index zero. This
is a consequence of the linear analysis in Section \ref{sec:lsa} and is
therefore the same regardless of the nonlinearity in
(\ref{eq:sh35}). Continuation of this localized state produces
characteristic snaking behavior [Fig.~\ref{fig:compabeta0}],
qualitatively similar to that present in the quadratic-cubic case
\cite{radlloyd,mccalla}. We therefore refer to this branch as the {\it spot}
branch. As one follows this branch, a half-wavelength is added
to the spot solution after every other saddle-node 
and the solution thereby grows in spatial extent until it
fills the domain, becoming a {\it target} state [Fig.~\ref{fig:solutions}(b)]. 
Using \pdep\ over the disk we computed in parallel the linear
stability of the solutions along this branch, with stable (unstable)
segments of the branch indicated by thick (thin) lines 
[Fig.~\ref{fig:compabeta0}(b)]. The result shows that most of the
localized states on this branch are unstable, except for short
segments after every fold on the left. Stable large amplitude domain-filling
{\it target} states acquire stability shortly after the leftmost fold.

The transitions just described, namely the continuous evolution from the primary bifurcation into a
spatially localized snaking structure, followed by a continuous transition into a domain-filling state,
is characteristic of systems with nonstandard boundary conditions, as explained in \cite{mercader};
in problems with Neumann or periodic boundary conditions, such as the \SH equation on a periodic interval,
the localized solutions appear in a secondary bifurcation from a periodic state that occurs at small
amplitude, and these localized states reconnect to the domain-filling state near the fold of the latter.
Thus the transition to and from localized structures takes place via bifurcations involving the periodic
state, and does not occur smoothly following a single branch. In the present problem this departure from
the periodic case is a consequence of the nonautonomous nature of the radial problem and not the boundary
conditions.

\subsection{Axisymmetric solutions on a large disk}

\begin{figure}
  \begin{center}
     \ig[width=8.6cm]{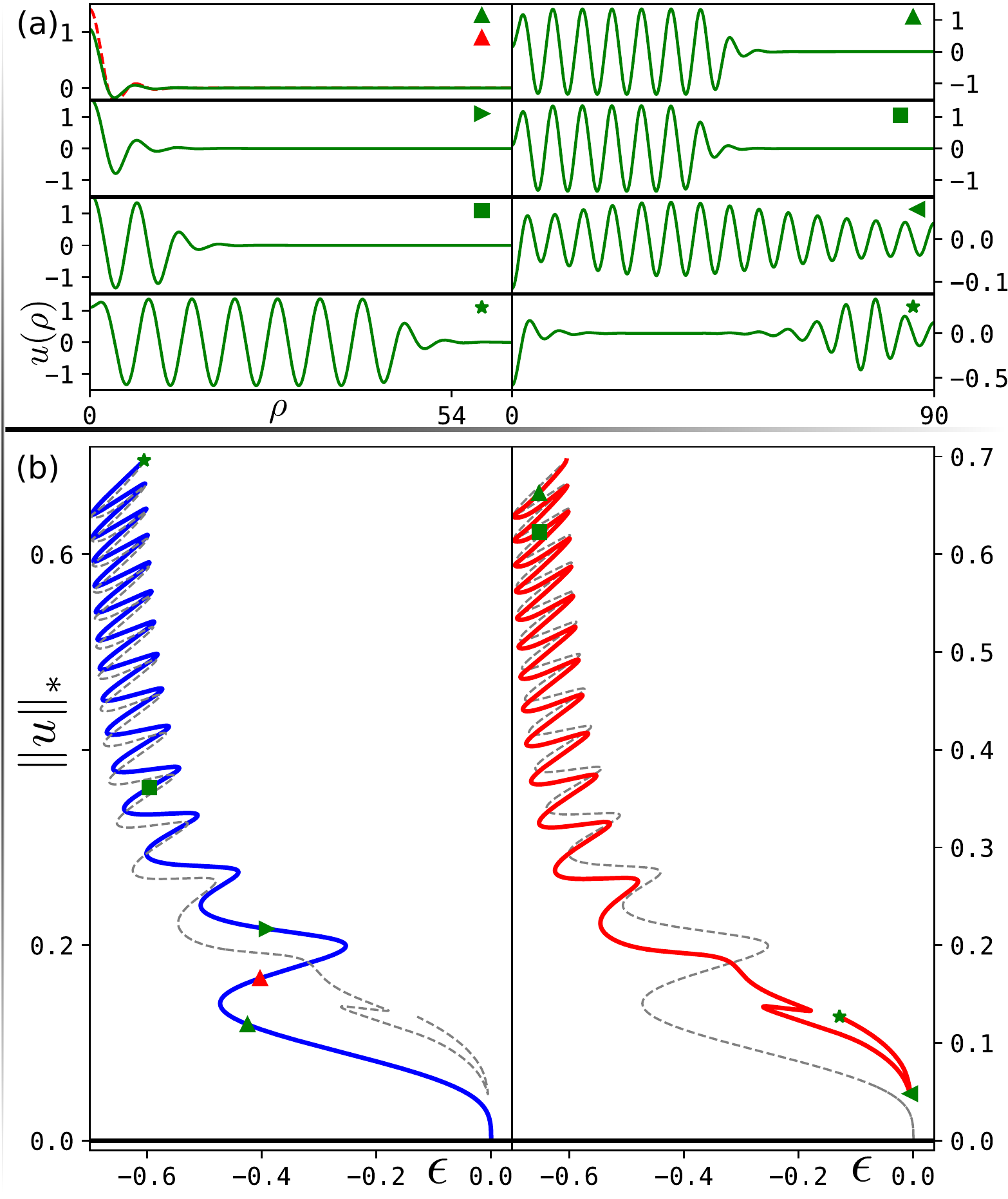}
  \end{center}
  \caption{Solution branch for the boundary value problem
    (\ref{eq:bvprad}) on a disk of radius $R=90$ when $\nu=2$, $q=1$ (b).
    {\it Spot} solutions are represented in blue (left) and evolve into {\it ring} solutions in red (right).
    The trivial solution $u=0$ is in black.
    Each segment of the bifurcation diagram is accompanied by representative
    solutions (a) shown at locations indicated by different symbols
    (triangles, squares and stars).}
  \label{fig:bvpl90}
\end{figure}

\begin{figure}
  \begin{center}
    \ig[width=6.8cm]{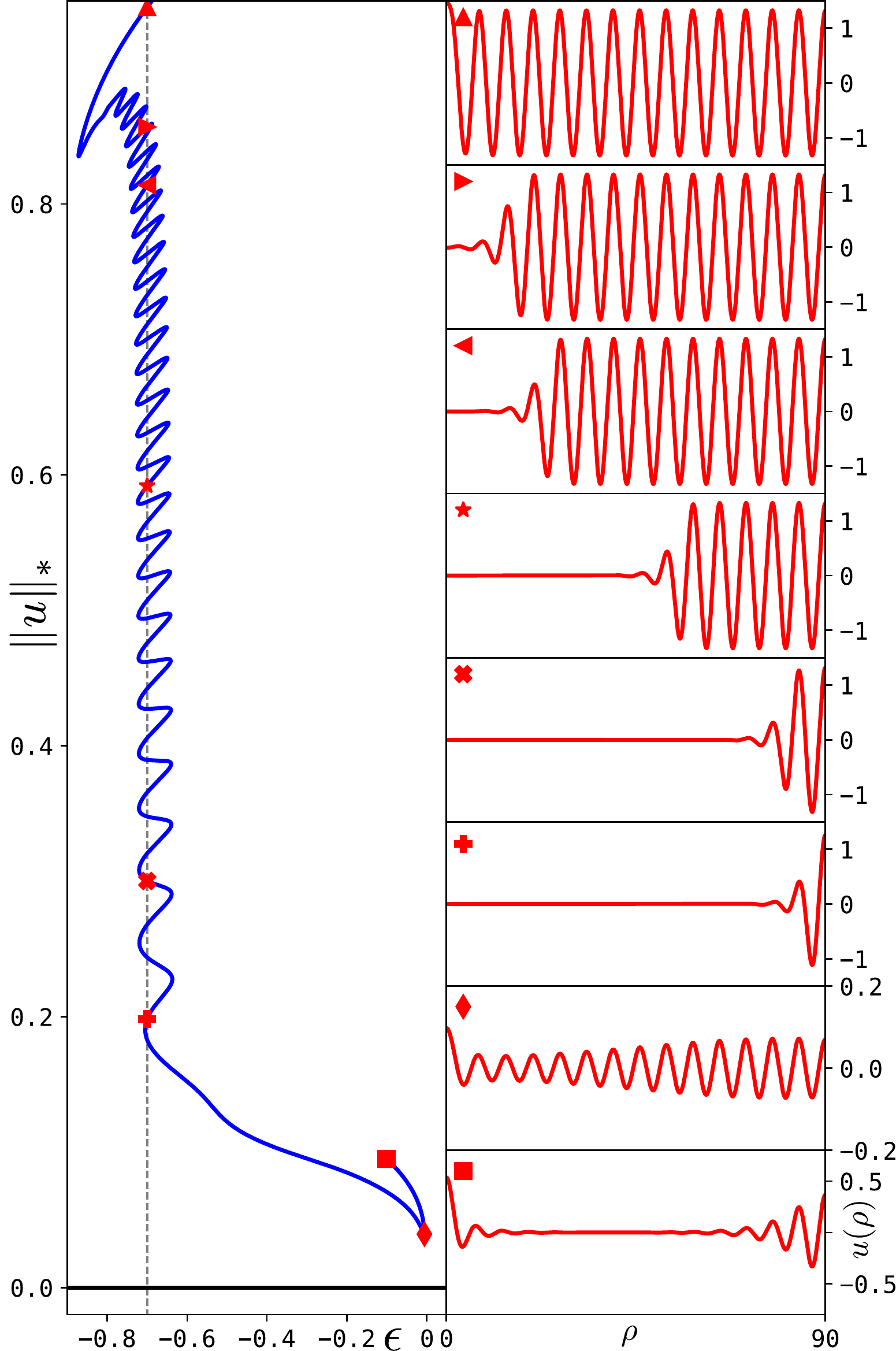}
  \end{center}
  \caption{Disconnected branch of solutions evolving from a target state for the boundary value problem
    (\ref{eq:bvprad}) when $R=90$, $\nu=2$, $q=1$ (left, blue curve). The trivial solution $u=0$ is in
    black. Different symbols (square, diamond, star, crosses and
    triangles) mark the location of representative solutions (most along the dashed line $\eps=-0.7$),
    illustrated on the right.}
  \label{fig:fromtargetaxi}
\end{figure}

Previous studies of axisymmetric solutions of the \SH
equation with a quadratic-cubic nonlinearity assumed an infinitely
large ($R\to \infty$) disk \cite{radlloyd,mccalla,bramburger}. In
order to illustrate the similarities and differences between these
studies and the cubic-quintic case studied here, we perform numerical
continuation of (\ref{eq:bvprad}) on a disk large enough ($R=90$) to admit
approximately 30 wavelengths across a diameter. The results are
summarized in Figs.~\ref{fig:bvpl90} and \ref{fig:fromtargetaxi}. 

As in the case with $R=14$ (Fig.~\ref{fig:compabeta0}), the branch of
localized states for $R=90$ bifurcates from $u=0$ at $\eps\approx 0$ as a spot solution,
undergoes a series of folds, adding a half-wavelength to the solution after
every other fold as in the unbounded case \cite{mccalla} or the 1D cubic-quintic \SH
equation \cite{burkesh351}.  However, in contrast with the case
$R=14$, the branch eventually ceases to add more oscillations and instead
starts to lose them.  Figure \ref{fig:bvpl90}(b) shows two
copies of this branch; the left and right panels highlight the portions
of the branch where the solution adds oscillations (blue) and loses oscillations (red). 

Figure \ref{fig:bvpl90}(b, left panel), shows that the process of adding oscillations
to the spot solution continues for a number of folds, but instead of connecting to the
target solution, there is now a transition where the solution changes from having a
maximum at the origin (spot) to a local minimum (ring) [green star in (a)]. Further
continuation of the branch [Fig.~\ref{fig:bvpl90}(b, right panel)] leads to a progressive
loss of oscillations by the same mechanism, thereby reversing the process described above.
This contraction process continues until the ring solution has only one maximum and one
minimum. Subsequently the solution branch connects to a right-most fold located at
$\eps \approx0$ (left-pointing triangle), corresponding to a small amplitude spatially
modulated periodic state. Further continuation of the solution past this
fold results in states with an additional ring-like structure near the wall at
$\rho=R$. This state, indicated by a star symbol in
Fig.~\ref{fig:bvpl90}(b, right panel), is thus a combination of a spot at the origin and a ring-like
structure along the wall. Like the states centered on $\rho=0$, these wall structures
are of two types, distinguished by whether they peak at the wall (a {\it wall spot})
or near the wall (a {\it wall ring}). Further continuation leads to further snaking
of these combination states, either gaining or losing maxima, and
eventually passing through a fold and turning into a yet different
combination state (not shown) in a process that does not appear to terminate.
The exact details depend strongly on the size of the
disk, much as found in the 1D problem with Robin boundary conditions \cite{houghton}
and have not been studied in detail.

\begin{figure}
  \begin{center}
    \includegraphics[width=8.6cm]{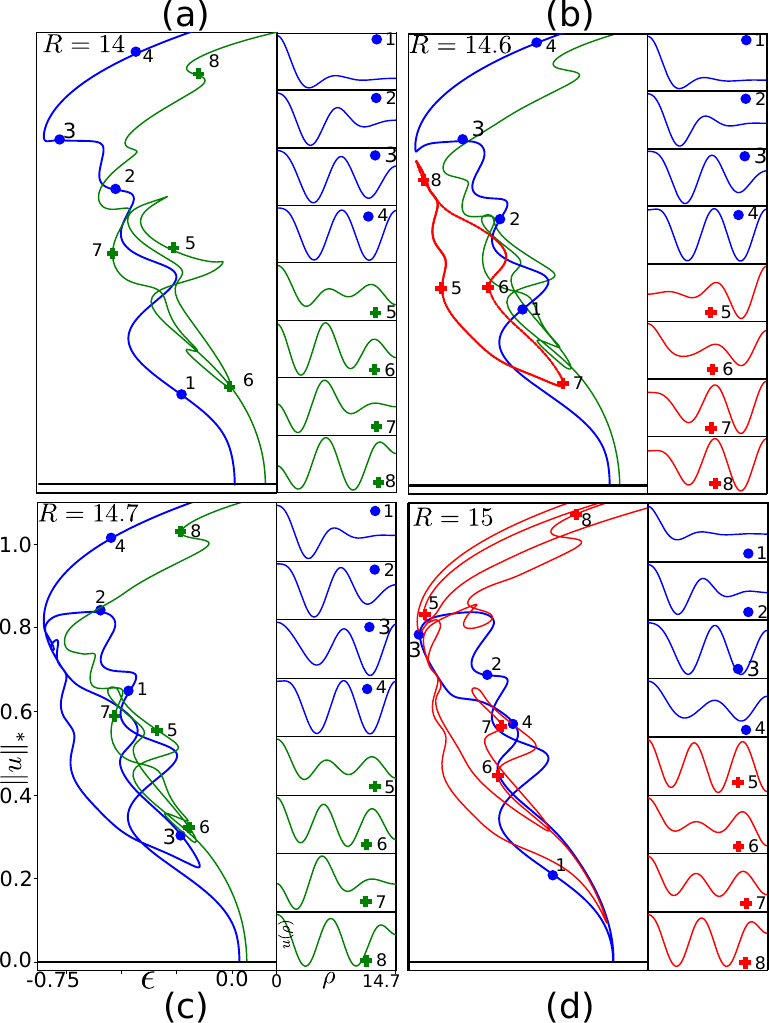}
  \end{center}
  \caption{Bifurcation diagram for axisymmetric states when (a) $R=14$, (b) $R=14.6$,
    (c) $R=14.7$, and (d) $R=15$ illustrating the pair of disconnections that are responsible
    for transforming the structure of (a) into (d). Sample radial profiles, at locations indicated by symbols
    (blue disks or red crosses), are included alongside.}
    \label{fig:discon}
  \end{figure}

\begin{figure}
  \begin{center}
    \includegraphics[width=8.6cm]{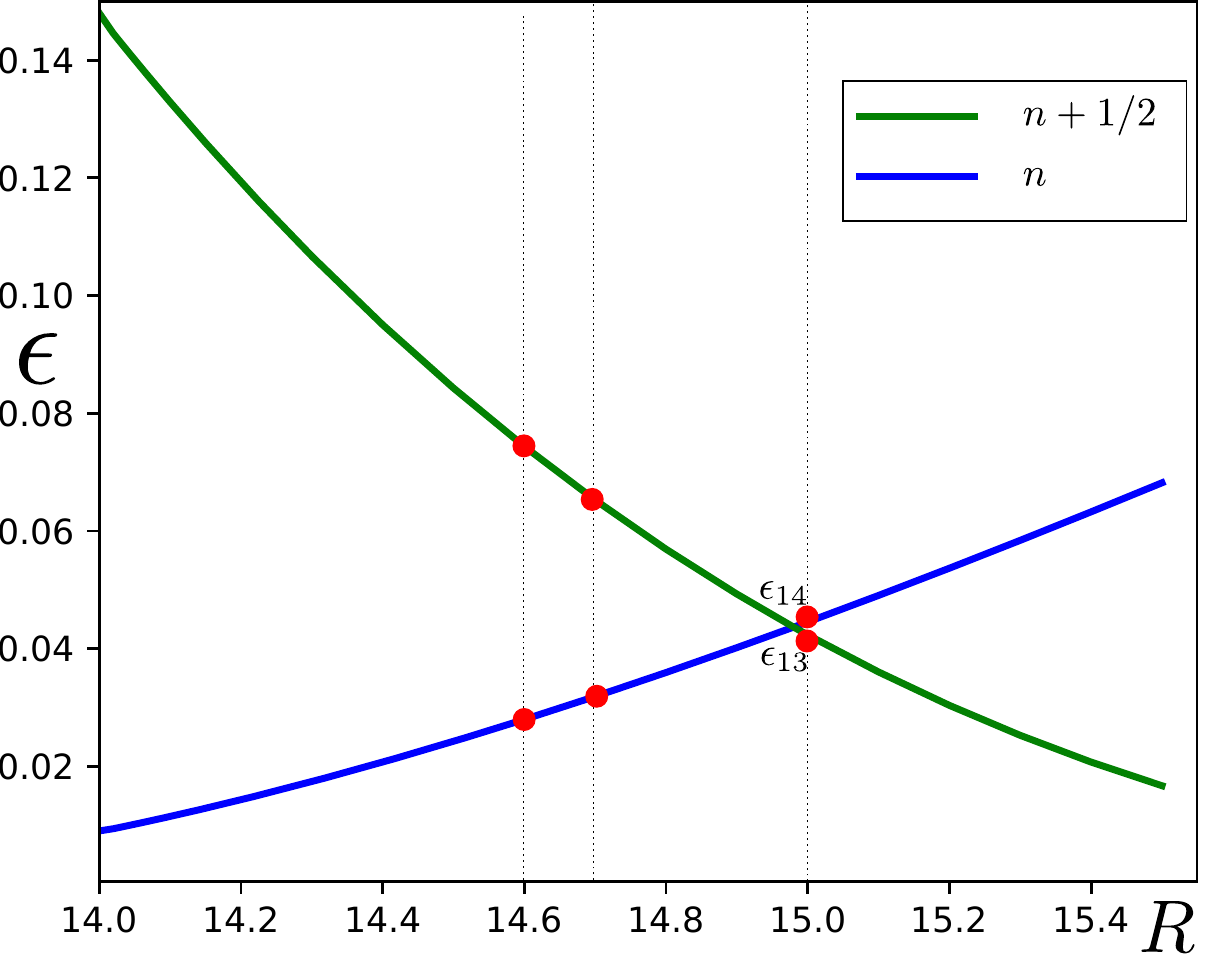}
  \end{center}
  \caption{Mode crossing in the linear stability problem (\ref{eq:sh35n0}) as a function of the radius $R$ when $q=1$.
    A marginal mode with $n$ wavelengths (blue) is superseded by a marginal mode with $n+(1/2)$ wavelengths (green) at
    $R\approx 14.98$. Here $n=2$. The eigenvalues $\eps_{13}$ and $\eps_{14}$ at $R=15$ are indicated with red dots as
    are the corresponding eigenvalues at $R=14.6$ and $14.7$, the radii employed in Fig.~\ref{fig:discon}.
    }
 \label{fig:linzeros}
  \end{figure}

\begin{figure*}
  \begin{center}
    \includegraphics[width=17.2cm]{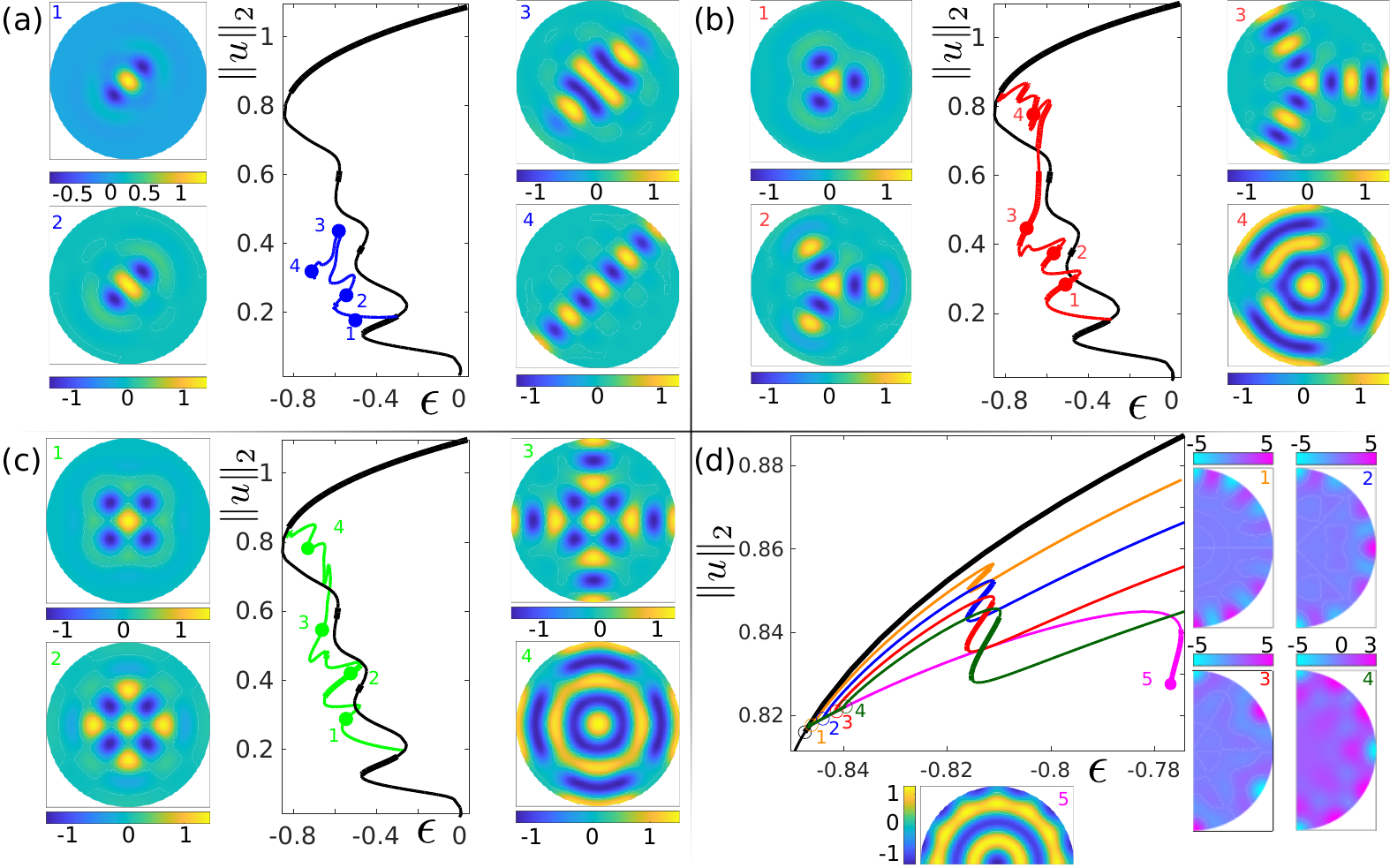}
  \end{center}
\vs{-5mm}
  \caption{Branches of multiarm solutions for $R=14$, $\nu=2$, $q=1$.
(a)--(c) $m=2,3,4$ branches bifurcating from an axisymmetric spot at low $\|u\|_2$. 
Stability with respect to perturbations on the full disk is indicated by thick line
segments. In each case, illustrative solutions are numbered and shown in corresponding insets. 
(d) Zoom of the bifurcation diagram with the $m=12$ crown branch (magenta) and the branches of 1-arm
(orange), 2-arm (blue), 3-arm (red), and 4-arm (green) states that bifurcate from it; the $m=3,4$-arm
states connect to the corresponding branches in panels (b) and (c). The four critical eigenfunctions
on the $m=12$ crown branch are shown alongside, together with the solution
profile at location 5 on this branch.  See Fig.~\ref{maf2c} for further 
continuation and sample profiles along the $m=1,2$ 
branches bifurcating from the first two bifurcation points on the crown branch. }
  \label{fig:multiarm}
  \end{figure*}

In contrast with the {\it spot} to {\it ring} transition described above,
the existence of the {\it target} solution is independent of the radius of
the disk. Figure \ref{fig:fromtargetaxi} shows a target solution for $R=90$
(see the up-pointing triangle in inset) and its connection to the {\it wall spot}
solutions (e.g., right-pointing triangle in inset). Further continuation of the
branch leads to wavelength loss, much as observed after the transition from spot
to ring [Fig.~\ref{fig:bvpl90}(b, right panel)]. The different insets illustrate
this process (left-pointing triangle, star, and crosses). The wall-spot solution
reaches a single maximum close to the prominent fold at $\eps\approx0$ (diamond
in inset) heralding a transition to a new combination state, consisting
of a wall spot at $\rho=R$ and a spot at the center of the domain, $\rho=0$.
As far as we can tell both branches (that bifurcating from $u=0$ and the
continuation of the target branch) execute similar repeated upward and downward
snaking but never  connect, i.e., the target state remains disconnected from $u=0$.

To understand this disconnection process we return in Fig.~\ref{fig:discon} to the
moderate $R$ case and discuss the transition from $R=14$ (direct connection of the
primary $u=0$ branch to the target) to $R=15$ (no such connection) in
greater detail. Specifically, the figure shows the bifurcation
diagrams for $R=14$, $R=14.6$, $R=14.7$ and $R=15$ together with
sample solutions. The plot at $R=14$ includes the second branch of
axisymmetric states (green curve) that bifurcates from $\eps\approx 0.148$,
in addition to the first branch (blue) that arises already at
$\eps_5$. At this value of $R$ these two branches are distinct
[Fig.~\ref{fig:discon}(a)] but this situation changes as $R$
increases. At $R=14.6$ [Fig.~\ref{fig:discon}(b)] the spot state is
still connected to the large amplitude target state although we are
beginning to see a cusp-like feature in place of the leftmost fold on the blue branch.
In addition, we have found a branch of axisymmetric states lying on an
isola (red curve). Such isolas of spatially extended states are present even in
the 1D cubic-quintic \SH equation \cite{kao}. This isola also possesses a cusp-like
feature in the vicinity of the cusp on the spot branch indicating that a codimension-two reconnection
is about to happen. Figure~\ref{fig:discon}(c) for $R=14.7$ describes the situation
after this reconnection. The spot branch (blue) remains connected to the target states but
now incorporates the isola states in the process. State 4 on the first target branch
differs from state 8 on the second target branch by approximately half a wavelength, as expected
of solutions connected via a fold.

A further transition occurs by $R=15$ [Fig.~\ref{fig:discon}(d)]. Figure~\ref{fig:discon}(c) shows that the leftmost
fold on the second spot branch (green) is approaching close to the first branch (blue) at $\eps\approx -0.76$ and amplitude
$\|u\|_*\approx0.76$, suggesting that a further reconnection takes place. This is indeed the case and Fig.~\ref{fig:discon}(d)
shows that as a result the blue branch becomes disconnected from the target states, and instead connects back to $u=0$
at $\eps_{14}$, i.e. at the origin of the second spot branch. The proximity of $\eps_{13}$ and $\eps_{14}$ at this value of
$R$ is responsible for the appearance of the cusp-like feature at $u=0$ in the blue branch seen in the figure. Moreover,
the target states on both the blue (profile 4) and green (profile 8) branches in Fig.~\ref{fig:discon}(c) are now connected.
These are part of a sequence of small and large amplitude states that are disconnected from $u=0$ and shown in red. These
may form via the incorporation of additional isolas into the branch. In fact, at $R=15$ the target states are part of a
heavily folded upper branch responsible for the appearance of additional target states at large amplitude.

Evidently, the transition that leads to the break up of the branch connecting the first spot branch to the
corresponding target states is exceedingly complex. In fact, we expect that the transition shown in Fig.~\ref{fig:discon}
is one of many such transitions. This is because the system has a preferred radial wavelength. As a result,
as $R$ increases, the system must repeatedly execute disconnections of this type, in order that the new lowest
radial mode is again able to connect to the target state with the correct number of wavelengths. These transitions
are triggered linearly when successive bifurcation points on $u=0$ pass through one another as $R$ increases
(Fig.~\ref{fig:linzeros}), and these are inevitably associated with additional nonlinear transitions of the type
shown in Fig.~\ref{fig:discon}. Similar behavior occurs in 1D as well \cite{bergeon}.

\section{Nonaxisymmetric solutions on a disk}\label{sec:nonaxi}

\subsection{Multiarm states}
\label{sec:mutiarm}
We now explore secondary bifurcations from axisymmetric states 
to nonaxisymmetric structures. For this 2D problem we use \pdep\ on the half disk of radius 
$R=14$ but compute the stability of the solutions on the
whole disk. The results are presented using the 2D $L^2$ norm
\huga{\label{u2def2} 
  \|u\|_2=\sqrt{\frac 1{|\Omega|}\int_\Omega u^2({\bf x})\dd {\bf x},}
  }
with $\eps$ as the continuation parameter unless otherwise stated.
The results are summarized in Fig.~\ref{fig:multiarm}. 

For $R=14$, the $m=2$ mode is the first nonaxisymmetric 
mode that bifurcates from the axisymmetric state with a spot
at the center of the domain [Fig.~\ref{fig:multiarm}(a)]. The bifurcation is 
characterized by the appearance of new maxima on opposite sides of the spot (state 1).
The resulting branch snakes, a behavior that is associated with the successive nucleation
of additional maxima at the tips of the resulting arms after every second fold (states 2 and 3),
in a process that resembles the growth of localized structures in the 1D cubic-quintic \SH equation.
However, in contrast to the 1D \SH equation, as this is taking place the structure broadens into a
{\it worm} \cite{avitabile} prior to reaching the boundary at $\rho=R$ and refocusing (state 4).
In addition, further continuation of the $m=2$ branch did not lead to lateral broadening of the arms
and the branch does not approach a domain-filling target state, in contrast to the cases $m=3,4$
discussed next. Instead the $m=2$ arms retract towards the boundary
forming a pair of spots at opposite ends of a diameter (not shown).

We next discuss the $m=3$ [panel (b)] and $m=4$ [panel (c)] branches. These consist of 3-arm and
4-arm states, respectively. Like the $m=2$ branch, 
these branches start at low norm on the branch of axisymmetric states (state 1), and grow by essentially
the same snaking process as the $m=2$ branch until they reach the boundary. However, at this point 
snaking gives way to a 'vertical' increase in the norm that is associated with rapid lateral expansion of the
arms. This type of growth is not associated with snaking because of the absence of pinning in this
direction \cite{avitabile} and takes place in the vicinity of the Maxwell point, i.e., the $\eps$-value
at which the energy $\mathcal{F}[u]$ of the target state vanishes. This lateral expansion generates states
that almost completely fill the domain, leaving 3 (resp. 4) radial gaps or holes. The snaking observed at
the top of this interval of rapid expansion is associated with the successive formation of complete rings
around the disk center, i.e., with the gradual retraction of the radial holes towards the boundary.

Surprisingly, we find that neither state connects to a domain-filling target
state. Instead, this connection takes place via a new, intermediate state we refer to
as a {\it crown} state. This state bifurcates from the target branch at the point where
the target branch stabilizes, and represents a near-axisymmetric state with an $m=12$ modulation
superposed on the stripe adjacent to the wall (magenta branch in panel (d) and sample plot 5).
The right column in panel (d) shows the first 4 bifurcation directions along the magenta branch.
These correspond to the appearance of 1, 2, 3 and 4 equispaced holes 
along the wall, and we find that the $m=3,4$ branches in panels (b,c) terminate on the
$m=12$ crown branch at the corresponding $m=3,4$ bifurcations (red and green branches).
Thus, contrary to expectation, none of these states connect directly to the target state.

The 1-arm (orange) and 2-arm (blue) hole states that bifurcate from the magenta branch in panel (d)
have been continued away from the crown branch and the results are illustrated in Fig.~\ref{maf2c}.
Both branches extend to low norm and do so via rapid broadening of the hole that turns the hole into
a 1-arm (resp. 2-arm) state. However, neither branch connects to the branch of axisymmetric spots.
Instead, the 1-arm state gradually retracts towards the boundary via snaking and turns into a
wall-attached spot (state 4 in panel (a)). The 2-arm state likewise shrinks into a pair of spots
on the boundary (state 4 in panel (b)). In both cases these spots subsequently regrow new spots
towards the interior, but no longer along the radius (not shown).
\begin{figure}
  \begin{center}
    \includegraphics[width=8.6cm]{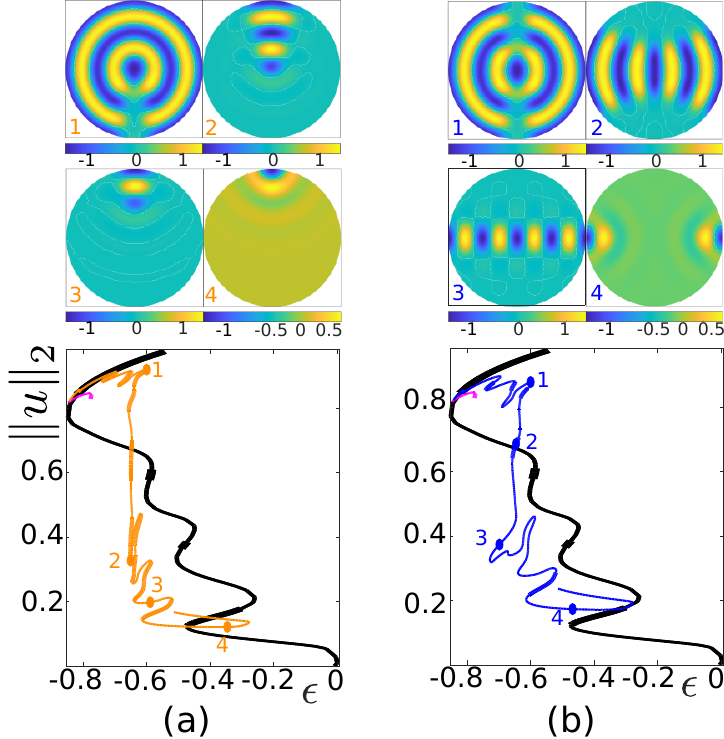}
  \end{center}
\vs{-5mm}
  \caption{Continuation of Fig.~\ref{fig:multiarm} showing the $m=1,2$ branches 
    from their origin on the $m=12$ crown branch. Stability with respect to perturbations
    on the full disk is indicated by thick line segments. \label{maf2c}}
  \end{figure}

In summary, for $m=3,4$ we have a connection between the radial spot 
at low norm and the $m=12$ crown branch very close to the top 
left fold on the target branch. For $m=2$ we have two disconnected
branches of 2-arm states,
one emerging from a small amplitude spot, and the other from the $m=12$ crown branch
at the top, both exhibiting very similar 2-arm states in between, for $\|u\|_2\approx 0.4$, say, but 
the two branches do not connect. For $m=1$ we only have one branch, with 
no bifurcation to a 1-arm state from an axisymmetric spot at low 
$\|u\|_2$. Remarkably, all of these $m$-arm branches include some
stable states. 

\begin{figure}
    \begin{center}
       \ig[width=8.6cm]{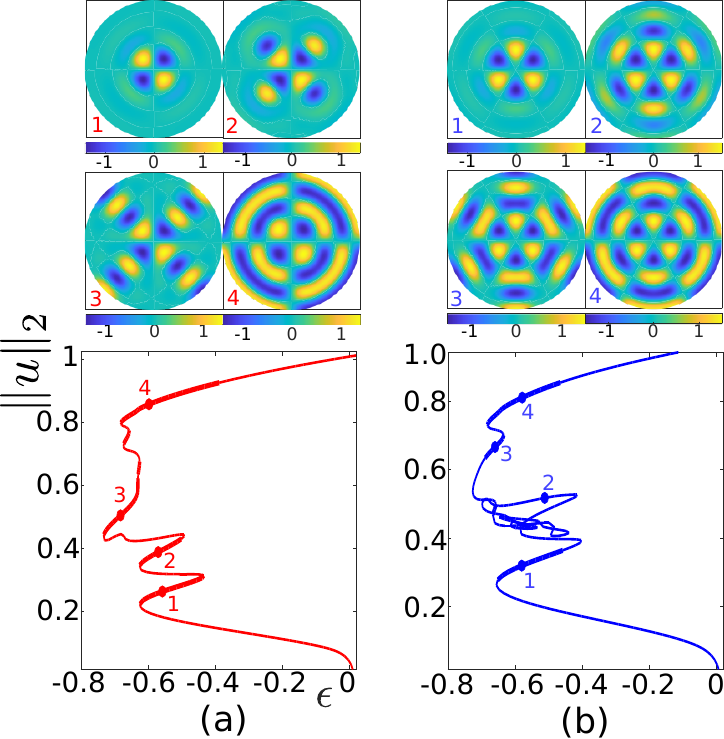}
    \end{center}
    \caption{Branches of solutions with the symmetry $D_m^-$ emerging from bifurcations of the trivial
      state $u=0$. Panels (a) and (b) show the $4^-$ and $6^-$-arm states, respectively. The bottom
      panels show the corresponding bifurcation diagrams in red and blue. Sample states at locations indicated
      by filled dots are shown in the top panels. Stability with respect to perturbations on the full disk is
      indicated by thick line segments.}
    \label{fig:minus1}
    \end{figure}

The $m=1,2,3,4$-arm states computed above are invariant under rotations 
$\phi\to\phi+2\pi/m$ and reflections across a suitable line. 
These operations generate the group $D_m$ and our continuation
procedure respects this symmetry, unless tertiary bifurcations take place that break it.
$D_m$-symmetric states may also bifurcate directly from the trivial state, see, for instance, the 
2nd, 3rd and 6th mode in Fig.~\ref{fig:linpred}(a). In addition, 
branches of $D_m^-$-symmetric states are also present; these states 
change sign upon rotation by $2\pi/m$. In Figs.~\ref{fig:minus1}(a) and (b) 
we show two such states, with symmetry $D_4^-$ and $D_6^-$, respectively.
These states are $D_2$ and $D_3$-invariant but not $D_4$ and $D_6$-invariant. 

Like the multiarm solutions of Fig.~\ref{fig:multiarm} and 
Fig.~\ref{maf2c}, the $D_m^-$ states start as localized structures near
the center of the domain but develop $m$ arms as one proceeds to larger amplitude which broaden into
extended $D_m^-$ target states after reaching the wall (insets 1-3 in each case). Because of the $D_m^-$
symmetry these states fill the domain with a target-like structure consisting of
$2\pi/m$ slices of the disk with the same pattern but alternating sign (see inset 4 in each case). 
The $D_m^-$ symmetry is present in the problem because of the $u\to -u$ symmetry of the cubic-quintic \SH
equation. Mathematically, the same symmetry is present in porous media convection and it therefore comes as
no surprise that states similar to the $m=4$ states in Fig.~\ref{fig:multiarm}(c) were found in numerical
continuation of the relevant equations of fluid dynamics on a periodic domain with a square cross-section
\cite{lojacono1,lojacono2}. In particular, in this problem one finds both $D_4$ and $D_4^-$-symmetric
structures. However, because of the periodicity of the spatial domain, these states bifurcate from a
branch of periodic structures rather than directly from the trivial state.

\subsection{Daisy states and their bifurcations}
\label{sec:daisy}

Our next family of nonaxisymmetric states is characterized by high azimuthal wave number. 
For this reason these states are localized towards the periphery of the domain, i.e.~they are wall modes.  
Following \cite{leberre} we refer to them as daisy states.
Figure~\ref{fig:solutions}(g) shows an example 
for $R=14$ and $\nu=1.5$. The daisies are periodic in the
angle $\phi$ and have symmetry $D_m$, $m\gg1$. 
Consequently they are easily computed on smaller sectorial domains
with opening angle $2\pi/m$ or even $\pi/m$; in the latter case the solution will have the property
$u(\rho,\pi/m)=-u(\rho,0)$ for all $\rho$, with $u_{\phi}(\rho,\pi/m)=u_{\phi}(\rho,0)=0$, modulo rotations
of the state. 

The blue branch in Fig.~\ref{fig:daisy} represents the computed daisy states, 
with sample states at locations \textbf{wu} and \textbf{wd} shown in the top panels. 
The daisy states bifurcate subcritically from the trivial state $u=0$ at $\eps_3$ and acquire linear
stability beyond the left fold of the blue branch, although they lose it again at yet larger amplitude,
see Fig.~\ref{fig:stripdaisy} and Fig.~\ref{wf1} below for details. As before, the notion of stability
always refers to stability with respect to perturbations on the whole disk, even for solutions computed
on smaller sectors. Since the daisies represent subcritical periodic states, we expect to find, by
analogy with the 1D cubic-quintic \SH equation \cite{burkesh351}, localized wall states resembling
a partially plucked daisy in appropriate parameter regimes. We further expect snaking branches 
of such plucked daisies where petals are added/removed after every other 
fold. To obtain these states we start with a moderate subcriticality,
$\nu=1.4$, and discuss how the scenario changes for 
stronger subcriticality, namely $\nu=1.5$. In Fig.~\ref{fig:daisy}
we show branches of localized daisies of odd (green) and even (red) parity;
sample states are shown in the panels on the right of the figure, together with 1D profiles along
their periphery. These illustrate the analogy between
these states and the 1D \SH scenario of \cite{burkesh351}. As in the 1D \SH case, the
odd and even branches of localized daisy states are connected via rung states, shown in black and 
in the inset labeled \textbf{r}. There are in fact four such branches owing
to the symmetry $u\to -u$. These localized daisies
bifurcate from the daisy branch at small amplitude (bottom right corner of the bifurcation diagram
in Fig.~\ref{fig:daisy}), undergo snaking whereby the solution adds half a wavelength at either end
after every left fold, eventually leaving just one plucked petal of the daisy, before terminating on
the complete daisy state near its fold, likely via Eckhaus instability, cf.~\cite{bergeon}. 
In addition to the even and odd branches of localized states shown in
Fig.~\ref{fig:daisy}, we also found 2-pulse localized daisies (magenta branch 
in Fig.~\ref{fig:stripdaisy}(a) as well as 3- or more--pulse states 
(not shown) all of which that behave similarly to the 1-pulse case.

The above 2-pulse states lie on a continuous branch that snakes
just like the 1-pulse branch but bifurcates from the daisy state at larger amplitude
than the 1-pulse branch, and likewise terminates on this branch farther from its fold.
This is because the computed 2-pulse states are equidistant and so behave like
1-pulse states on the half-domain. We conjecture that non-equispaced 2-pulse
states lie on isolas, like the corresponding states in the 1D \SH equation \cite{burkesh2009},
but have not computed such states.

\input{df1}
\input{df2}
\input{df3}

It turns out that the daisy snaking scenario in Fig.~\ref{fig:daisy} 
depends rather strongly on $\nu$. In Fig.~\ref{fig:stripdaisy}(a) we again show 
the daisy branch at $\nu=1.4$ in blue, the branch of even 1-pulse states in green 
and even 2-pulse states in magenta, and number representative folds 1--4. 
Panels (b) and (c) show the continuation of these folds 
in the $(\nu,\eps)$ plane. In the 2-pulse case the fold
extends beyond $\nu=1.5$ but retreats at larger $\nu$, 
yielding two possible solutions at $\nu=1.5$ [magenta
triangles, panel (d)]. In the 1-pulse case, folds 1 and 3
exist at $\nu=1.5$ (and beyond), whereas fold 2 [in blue, panel (b)]
turns back before $\nu=1.5$.

   Using the solutions at $\nu=1.5$ [vertical dashed line in Figs.~\ref{fig:stripdaisy}(b) and (c)] as
   starting points, we return in Fig.~\ref{fig:nu15} to one-parameter
   continuation. Keeping $\nu=1.5$ fixed and continuing the 2-pulse state both up and down in amplitude
   (green and brown curves, respectively) yields the complete snaking branch of even states of this type.
   [Fig.~\ref{fig:nu15}(b)]. This branch connects with the daisy branch at the top and
   bottom. We can also perform one-parameter continuation of 
   the 2-pulse state in $\nu$, leading to the isola shown
   in Fig.~\ref{fig:nu15}(a). Profiles 1-2 show that the states
   on this isola correspond to a localized daisy whose petals at $\phi\approx 0$ are beginning to expand
   into the interior of the disk.

   Since fold 2 of the 1-pulse branch does not exist at
   $\nu=1.5$ [Fig.~\ref{fig:stripdaisy}(a) and (b)], we
   expect to see a disconnection of folds 1 and 3 when performing continuation in
   $\eps$. This is illustrated in Figs.~\ref{fig:nu15}(c) and (d). 
   The upper fold 3 from Fig.~\ref{fig:stripdaisy}(a) connects to
   the daisy branch near its fold [green segment in
   Fig.~\ref{fig:nu15}(d)], and continuation in the opposite
   direction yields snake-like behavior [brown segment in Fig.~\ref{fig:nu15}(d)]
   until some of the petals of the solution grow towards the interior
   of the disk, forming worm-like structures.
   Similarly, the lower fold 1 from
   Fig.~\ref{fig:stripdaisy}(a) connects to the daisy branch near the
   bottom, and continuation in the opposite direction also leads to
   snaking followed by the development of worm-like structures [Fig.~\ref{fig:nu15}(c)]. In
   both cases the continuation was stopped at an arbitrary point for the sake of clarity. 
   The 2-pulse snaking branch likewise breaks at $\nu\approx 1.535$ when fold 4 in
   Fig.~\ref{fig:stripdaisy}(a) disappears [Fig.~\ref{fig:stripdaisy}(c)].
   
In contrast with the 1D case, here the extra degree of freedom
in the radial direction leads to fat wall-attached worms
confined to part of the full disk. As a result, the branch of localized
daisy states is no longer able to terminate on the daisy branch and we have
been unable to determine its ultimate fate. We note that fat worm-like
states are well-known solutions of the \SH equation in the
plane, and these may be held together both by curvature of the
boundary and by pinning due a wave number gradient normal to the
boundary \cite{avitabile}; in other cases these states are unstable
leading to temporal growth \cite{lloyd2019}. We associate
the appearance of the fat worms with the initiation of the filling
transition, whereby the disk gradually fills with an extended stripe
state as parameters change. Evidently this transition occurs as a
result of the advance of the front connecting the wall-attached stripe
state to the trivial state, towards the origin of the disk. How the
location and shape of this front depend on parameters is an important
question; dynamical integration suggests that outside the snaking
region the worm state either collapses (to the left of the snaking
region) or else grows dynamically leading to dynamic filling of the
domain interior (to the right of the snaking region). 
See Sec.~\ref{sec:otherpatterns} for further discussion.

\section{Energy of solutions}
\label{sec:energy}

\begin{figure}
    \begin{center}
      \ig[width=8.6cm]{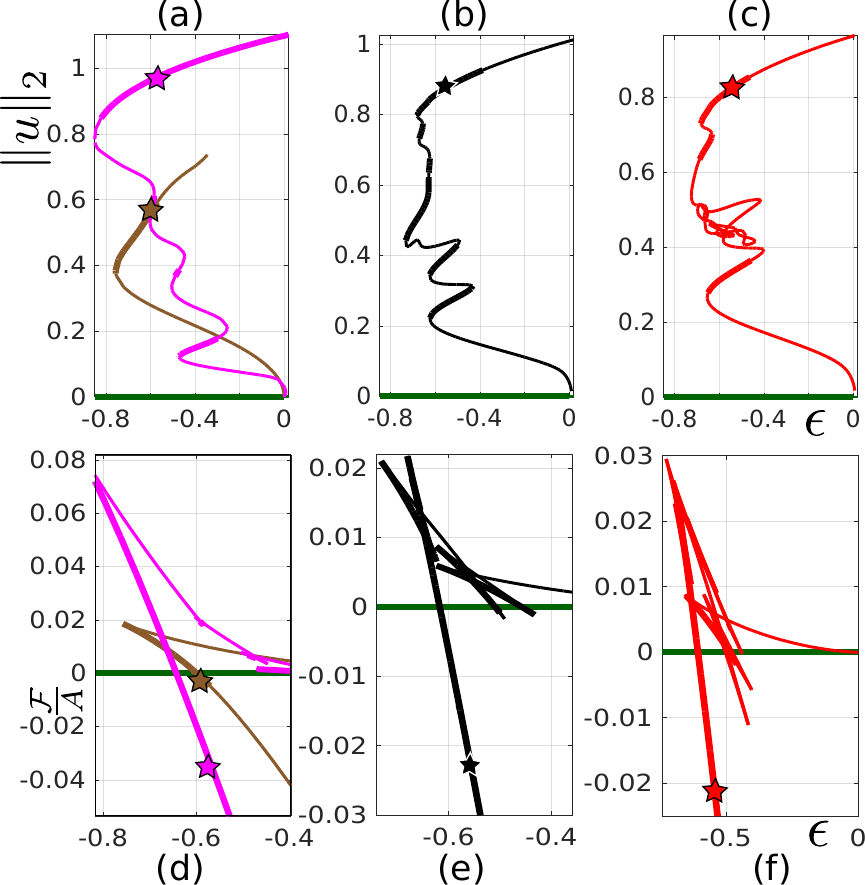}
    \end{center}
        \caption{Solutions for $\nu=2$ as a function of $\eps$ shown in terms of (a)--(c)
          their norm $\|u\|_2$ (\ref{u2def}) or (d)--(f) their energy (\ref{eq:lyapfunc}) per unit area.
          The trivial branch is in green. Panel (a) compares the axisymmetric (magenta) and daisy
          (brown) states.
          Panels (b) and (c) show the $4^-$-arm (black) and $6^-$-arm (red) branches, respectively.
          In each case, sample points are marked with a star to illustrate the mapping between the norm
          and the energy $\mathcal{F}/A$. All computations are for a disk with $R=14$, $q=1$.}
    \label{fig:ennu2}
  \end{figure}

\begin{figure}
    \begin{center}
       \ig[width=8.6cm]{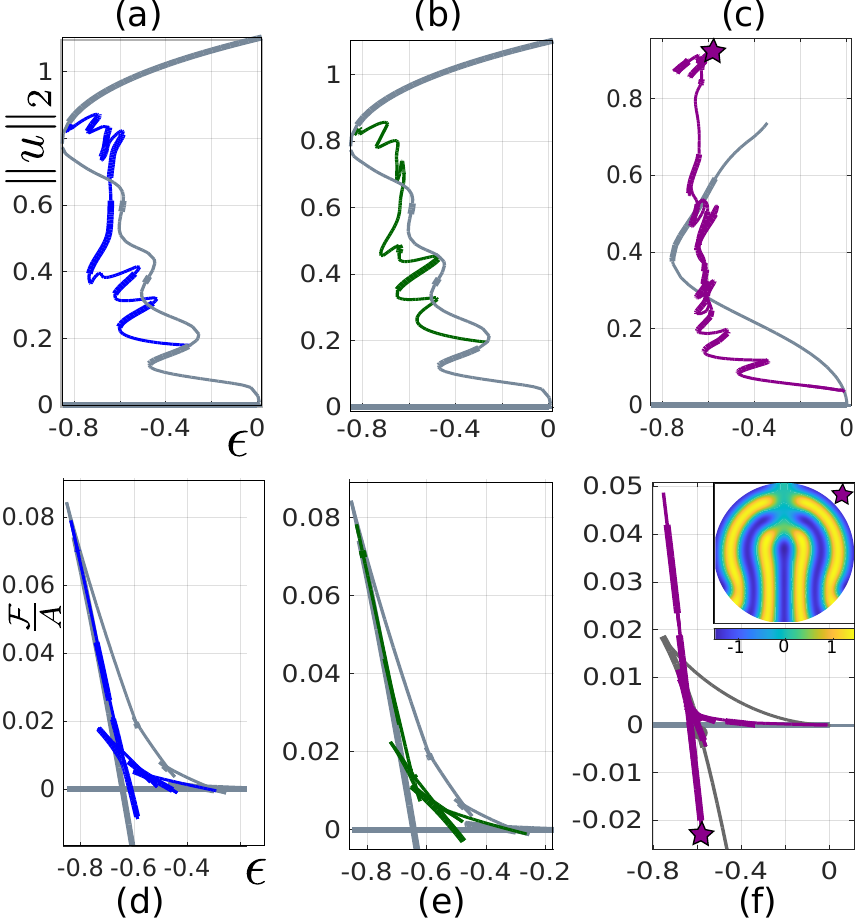}
    \end{center}
    \caption{Secondary bifurcations from (a), (b) the axisymmetric branches and (c)
      the daisy branch when $\nu=2$, using the same convention as in
      Fig.~\ref{fig:ennu2}. Blue and green branches represent the $3^+$ and $4^+$-arm solutions,
      respectively. Panel (c) shows the branch of localized daisy states (purple)
      bifurcating from the daisy branch (gray). The branch initially snakes (localized daisy states) but
      subsequent continuation leads to a worm-like state and then to
      an $\Omega$-shaped state (inset) as the stripes extend far into the domain (star).}
    \label{fig:ennu21}
  \end{figure}

\begin{figure}
    \begin{center}
       \ig[width=8.6cm]{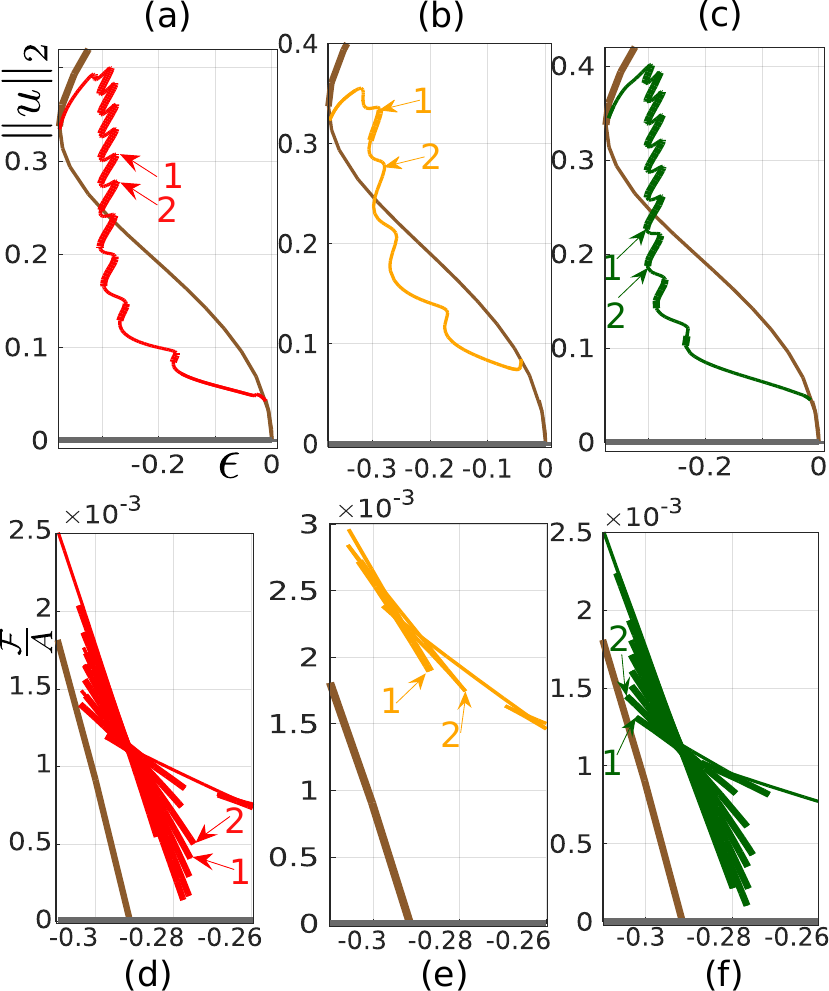}
    \end{center}
\caption{Solutions for $\nu=1.4$ using the same convention as in
  Fig.~\ref{fig:ennu2}. The daisy branch is shown in
  brown with branches of even [red, panel (a)], odd [green, panel (c)] and equispaced 2-pulse
  [yellow, panel (b)] localized daisy solutions also shown. Representative
  folds of the localized branches are numbered to enable comparison with the energy plots in
  panels (d)--(f).}
    \label{fig:ennu14}
  \end{figure}

Since the \SH equation seeks to minimize the free energy (\ref{eq:lyapfunc}),
the global energy minimum is of particular interest. In this section we therefore
compare the energies of the previously obtained solutions, restricting the discussion
to the parameter regime $R=14$, $q=1$, and treat the cases $\nu=2$ (Figs.~\ref{fig:ennu2}
and \ref{fig:ennu21}) and $\nu=1.4$ (Fig.~\ref{fig:ennu14}) 
separately, aiming to identify the branches with the lowest energy.
In both cases we show the norm $\|u\|_2$ [Eq.~(\ref{u2def})] and the corresponding
energy $\mathcal{F}/A$ [Eq.~(\ref{eq:lyapfunc})] in the top and bottom panels,
respectively. In the energy representation the folds in the former become cusps,
making the branches hard to distinguish. Consequently, we have split the branches into different 
subfigures, and indicate some sample solutions (using a star or an arrow) in order to illustrate
the mapping between these two representations. In all cases the energy vanishes close to the
Maxwell point for the target state.
  
Figure \ref{fig:ennu2} shows four branches emerging from the trivial
state (green) when $\nu=2$. Panel (a) compares the axisymmetric (magenta) and
daisy (brown) states. For small values of $\eps$, the least energy state is
the trivial state $u=0$. At $\eps\approx -0.63$, the energy of the axisymmetric states crosses zero
becoming negative and the minimum energy state is a target state.
For larger values of $\eps$, the energy of the daisy states also
becomes negative but remains larger than that of the target states.

The $m^-$-arm branches emerge from the trivial branch as localized
solutions. Further continuation leads to a series of folds, filling
the domain with a multiarm extended pattern with the symmetry
$D_m^-$. Figure~\ref{fig:ennu2} shows the branches for $m=4$ and $m=6$
in black [panel (b)] and red [panel (c)], respectively. As the branch
undergoes the above-mentioned series of folds, the free energy
oscillates between adjacent cusps. Continuation of the branch in the direction of increasing norm
leads to increasing energy until the solution reaches the fold at the top
left of the bifurcation diagram in the top panels. Beyond this point, 
the energy decreases again and eventually becomes negative. This occurs
first for $m=4$ and then for $m=6$. In both cases the
energy is always greater than that of the target state at the same value of $\eps$.

Figure \ref{fig:ennu21} illustrates the energy of the secondary
branches when $\nu=2$, showing the branches of the $3^+$ and $4^+$-arm
solutions in panels (a) and (b) and the corresponding free energy in
panels (d) and (e). The latter resemble the plots in Fig.~\ref{fig:ennu2}.
Panel (c) shows the fate of a localized daisy solution, after it emerges
from the daisy branch. As discussed in Sec.~\ref{sec:daisy}, the localized daisy
states initially add new petals every other fold. In the present case the solution subsequently
develops into a 1-worm-like structure instead of adding more petals, with the stripes extending
farther anf farther into the domain, eventually generating an $\Omega$-shaped structure
[panels (c) and (f)], in contrast to what happens in Fig.~\ref{wf1}(b) below.
Figure \ref{fig:ennu21}(f) shows that as the structure expands, energy starts
to decrease and the $\Omega$ state (star, inset) that results has negative energy.
At this value of $\eps$, this is the solution with the smallest energy after the target state.

As shown in Sec.~\ref{sec:daisy}, when $\nu=1.4$, the localized 
daisy states instead exhibit snaking behavior analogous to the 1D \SH equation \cite{burkesh351},
and this also holds for the energy. 
In Fig.~\ref{fig:ennu14} we show the even (red), odd (green) and equispaced two-pulse
(yellow) branches of localized daisy states, with representative folds marked with
arrows and numbered. In all cases, localized solutions with higher 
$\|u\|_2$ norm have lower energy. The rung states are not shown. The
2-pulse states always have greater energy than the odd  and even single pulse states.

In summary, the energy provides the following description in terms of
the lowest energy states: for small values of
$\eps$, the branch with the smallest energy is the trivial
one. As $\eps$ increases, axisymmetric states become 
energetically more favorable. For yet larger values of $\eps$, the
energy of the $4^-$ and $6^-$ branches becomes negative as is the energy of the $\Omega$ state
in Fig.~\ref{fig:ennu21}(f) but these states never become global
energy minima. In general spatially extended states have lower energy than localized states. 

\section{Further patterns}  \label{sec:otherpatterns}
  
We now briefly discuss some patterns to the right of the snaking region of 
the localized daisy states, and also perform some direct numerical simulations (DNS). 
The loss of stability of the daisy branches with increasing $\eps$  
generically results in wall-attached worms, but the subsequent behavior
is sensitive to the value of $\nu$.  In Fig.~\ref{wf1} we 
again contrast the cases $\nu=1.4$ and $\nu=2$ when $R=14$. Panel (a), 
computed on the half disk for $\nu=1.4$, shows that 
the first bifurcating branch (orange) contains states 
similar to those in Fig.~\ref{fig:nu15}, namely a 
stable wall-attached 3-worm which subsequently turns into a mix of wall-spots 
and interior spots, but does not become stable again. The next branch (green) 
bifurcates very close to the first, starts as a wall-attached 4-worm, and 
turns into a mix of stripes perpendicular 
to the wall with spots in between. Stable solutions exist
up to $\eps\approx -0.1$. Solution 4 is the result of
DNS from state 3, just after its loss of stability, and
yields vertical stripes near the disk center. Essentially the same 
states are obtained on the full disk, where DNS from unstable 
initial conditions generates various combinations of stripes near the center,
depending on meshing details, or differences in initial perturbations. 

For $\nu=2$ [panel (b)], the first bifurcation is to a $Z_4$-symmetric
wall-attached 4-worm, which subsequently expands into the interior of the disk
in the form of regular stripes but then breaks up, due to overcrowding, into a
mix of spots and stripes, leaving a hole at the disk center. DNS from the
$Z_4$-symmetric state 3 yields state 4, where the $Z_4$ symmetry is lost when
stripes recombine to fill the interior. 

\input{wf1}

The target patterns and the $m$-arm states from Sec.~\ref{sec:axi} 
all consist of stripes parallel to the wall, while the daisies, localized daisies,
and patterns 4 from Fig.~\ref{wf1}(a) and (b) consist of stripes perpendicular 
to the wall. In contrast, the $\Omega$ pattern from Fig.~\ref{fig:ennu21}(f) 
has both. Evidently, away from the stability regions of the patterns discussed so far, 
we may expect patterns of spots and stripes with some stripes 
parallel and others perpendicular to the wall. 
One example, shown in Fig.~\ref{paf1}, is the result of DNS at $(\eps,\nu,q)=(-0.6,2,1)$ 
starting at $t=0$ with parallel stripes $(u_1(x,y),u_2(x,y))=(\cos(4\pi y/14),0)$ in terms
of the 2-component second order system \reff{shsys0}.
The solution quickly converges to (the stable) state 1.
On continuing the solution from location 1 to larger $\eps$ we find that the pattern remains stable 
for all $\eps$ values reached without a qualitative change of shape. Continuation 
to smaller $\eps$ yields the magenta branch, which initially 
shows some snaking whereby the top and bottom stripes disintegrate into 
spots. Shortly after state 3 the branch loses stability and never regains it; the 
disintegration (and partial recovery) of the stripes continues, however, but the 
branch apparently never connects to any of the states already discussed. 
Similar behavior is obtained for $\nu=1.4$ and $\eps$ in the range $-0.3$ to $-0.1$ (not shown). 

\input{paf1}

\section{Discussion}
  \label{sec:discussion}
  We have explored numerically the variety of states described by the prototypical subcritical
  \SH equation on a finite disk with Neumann boundary
  conditions. We have chosen the cubic-quintic
  case in order to eliminate the preference for hexagonal
  structures in, for example, the quadratic-cubic case. The key point
  about the \SH equation is its characteristic scale $2\pi/q$. In the quadratic-cubic case, hexagonal
  structures on this scale are more easily accommodated within a
  disk of radius $R$ than the stripes that are preferred in the
  cubic-quintic case. Thus the latter is more interesting 
  from a physics point of view, and the number of wavelengths $2\pi/q$
  that can be accommodated along the diameter or the circumference of
  the disk becomes a key parameter. Throughout we focused on steady
  states since all time-asymptotic states of the model are necessarily
  steady.

  Because the primary bifurcation in our model is subcritical we
  expect both spatially extended structures and spatially localized
  structures. We focused on relatively small disks but even these already allow
  a large variety of (stable) patterns in the subcritical regime. Spatially
  extended states include target patterns consisting of concentric stripes parallel
  to the boundary and worm states with stripes normal to the boundary, while the localized
  structures can be divided into two types: those localized at the center of the domain and those 
  localized at the boundary. States exhibiting some stripes that are normal to the boundary and others
  that are parallel to it, such as the $\Om$ pattern in Fig.~\ref{fig:ennu21}(f), have also been obtained.
  However, neither PanAm-type patterns with convex stripes whose curvature is not imposed by the boundary
  nor spirals \cite{plapp} were found, suggesting that such states require large scale flows
  for their maintenance. Such flows are of course absent from the model studied here.

  For axisymmetric states we also considered the case of large $R$. The states then either have a
  monotonically decaying envelope and are then referred to as {\it spots}, or one
  that peaks away from the center or boundary, in which case we refer
  to them as {\it rings}. We found that in relatively small disks the central spot
  exhibited typical snaking behavior as it grew in extent, eventually
  filling the domain and becoming a target state. In contrast, in
  larger disks the spot state undergoes a transition to a
  domain-filling ring-like state that subsequently breaks up into a
  pair of ring structures, one near the center and one near the
  outer boundary. As one follows the solution branch further the
  connection between these states repeatedly forms and breaks, leading
  to exceedingly complex behavior of the solution branch.

  The axisymmetric states are subject to secondary
  symmetry-breaking bifurcations. We explored
  these on a relatively small disk, and identified secondary states with $D_2$, $D_3$ and
  $D_4$ symmetry, representing states with 2, 3 and 4 arms that
  gradually extend in length while remaining laterally localized. Once
  the arms reach the boundary, they begin to spread laterally, 
  terminating on a 'crown' state that bifurcates from a domain-filling target state close
  to its fold [Fig.~\ref{fig:multiarm}(d)]. 
  This is not the case, however, for primary $D_m^-$-symmetric states, for
  which rotations by $2\pi/m$ are equivalent to changing the sign of
  $u$ (Fig.~\ref{fig:minus1}). These states exhibit similar growth behavior but cannot connect 
  to a target state.

  In addition we also studied subcritical azimuthally periodic wall
  states (daisies), and showed that these were accompanied
  by azimuthally localized daisy states. 
  For weak to moderate subcriticality ($\nu<1.5$, say), these
  localized states grow in azimuthal extent in the same manner as
  localized states in the 1D cubic-quintic \SH equation,
  but for larger subcriticality we found that they instead expand into the
  interior, forming wall-attached worm-like states similar to those
  present in the plane \cite{avitabile,lloyd2019}.

  It is significant, though not altogether surprising, that the
  structures we identified in this simple model problem resemble
  similar structures observed in fluid flows, combustion, laser
  physics, and indeed other spatially confined systems. Convection in
  a vertical cylinder provides the closest realization of these states
  despite the absence of subcriticality. Numerical continuation 
  studies of this system \cite{tuckerman1,tuckerman2} in cylinders with
  moderately small aspect ratio identified not
  only target states but also nonaxisymmetric states with the symmetry
  $D_2$, $D_2^-$, $D_3$, $D_4$, $D_4^-$ and $D_6^-$. As in our case, some of these appear
  through a primary instability of the conduction state, while others
  appear through secondary bifurcations of axisymmetric states. In a
  similar vein, existing studies of porous medium binary fluid
  convection on a periodic domain with a square cross-section
  identified four-armed states with both $D_4$ and $D_4^-$ symmetry
  \cite{lojacono1,lojacono2} and studied their snaking behavior as the
  arms grew in extent, ultimately interacting with their images.

Despite the wealth of new phenomena described here, 
a number of critical questions remain. Even in our stripped down problem 
it proved impossible to follow many of the solution branches all the way,
and in larger domains it remains unclear whether the localized structures ever connect 
to a domain-filling state. The domain-filling transition of the wall state likewise
remains to be fully characterized. Much of the interesting behavior of this system
can be traced to the competition between stripes parallel to the wall or perpendicular to it.
This competition is sensitive to both the domain radius $R$ and the subcriticality parameter $\nu$,
and appears responsible for absence of a direct connection between the $D_m$-symmetric $m$-arm states
and axisymmetric target states. Nevertheless, this study should serve as a useful guide to 
subcritical pattern formation in a bounded 2D 
domain beyond the standard case of squares and rectangles with 
Neumann or periodic boundary conditions. 

\vspace{.3cm}

\appendix*

\section{The \pdep\ implementation}
 \label{sec:app}
In its standard setting, the \mlab\ package \pdep\ \cite{pde2path,p2phome,
hannesbook} uses the finite element method (FEM) to spatially 
discretize systems of second order PDEs, and combines this with 
a variety of numerical continuation and bifurcation algorithms, 
including some simple DNS. 
We therefore rewrite the 4th order \SH equation \reff{eq:sh35} as a 
parabolic--elliptic system (for simplicity setting $q=1$),
\huga{\label{shsys0}
M_d\pa_t \bpm u_1\\ u_2\epm{=}
\bpm -\Delta u_2{-}2u_2-(1{-}\eps)u_1{+}f(u_1)\\-\Delta u_1+u_2\epm, 
}
with a singular dynamical mass matrix $M_d=\bsmm 1&0\\0&0\esmm$, 
$f(u_1)=\nu u_1^3-u_1^5$, and 
Neumann BCs for $u_1$ and $u_2$. See, e.g., \cite[Remark 8.1]{hannesbook} 
for the equivalence of \reff{eq:sh35} and \reff{shsys0} over convex 
Lipschitz domains, or general domains with a smooth boundary. 
For the patterns studied here, it turns out that 
careful meshing \cite[\S4.1.1]{hannesbook} is crucial to maintain 
symmetry of the solution branches, 
i.e., to mitigate branch jumping, and throughout this work we used  
axisymmetric meshes. Additionally, instead of the standard 
piecewise linear FEM we chose 6-node triangles \cite[\S5.1]{poz14}, 
i.e., piecewise quadratic elements. With these, a typical discretization 
of the $R=14$ disk uses about 17 000 nodes, yielding a total of 
34 000 degrees of freedom for \reff{shsys0}. 

The basic \pdep\ implementation can be found in \cite{tuto},
where the main script produces the radial, daisy and localized daisy branches.
Additionally, movies illustrating the solutions along the branches
explored in this paper are included in the SI.


%% file: lpf2_ek.tex
 \begin{figure}[ht]
   \begin{center}
\includegraphics[width=8.6cm]{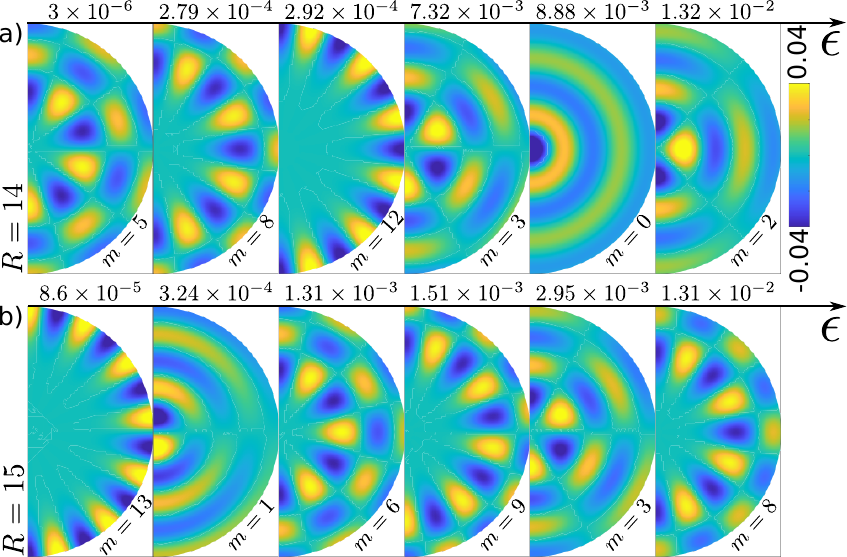}
\vs{-2mm}
    \end{center}
    \caption{The first six eigenfunctions on the half-disk 
predicted by linear stability analysis of the $u=0$ state when 
(a) $R=14$ and (b) $R=15$. The eigenfunctions are sorted according to the
eigenvalues $\epsilon$. In each case, $\epsilon$ and the corresponding
azimuthal number $m$ are specified.}
    \label{fig:linpred}
    \end{figure}
  

%% file: df1.tex
 \begin{figure}
    \begin{center}
      \ig[width=8.6cm]{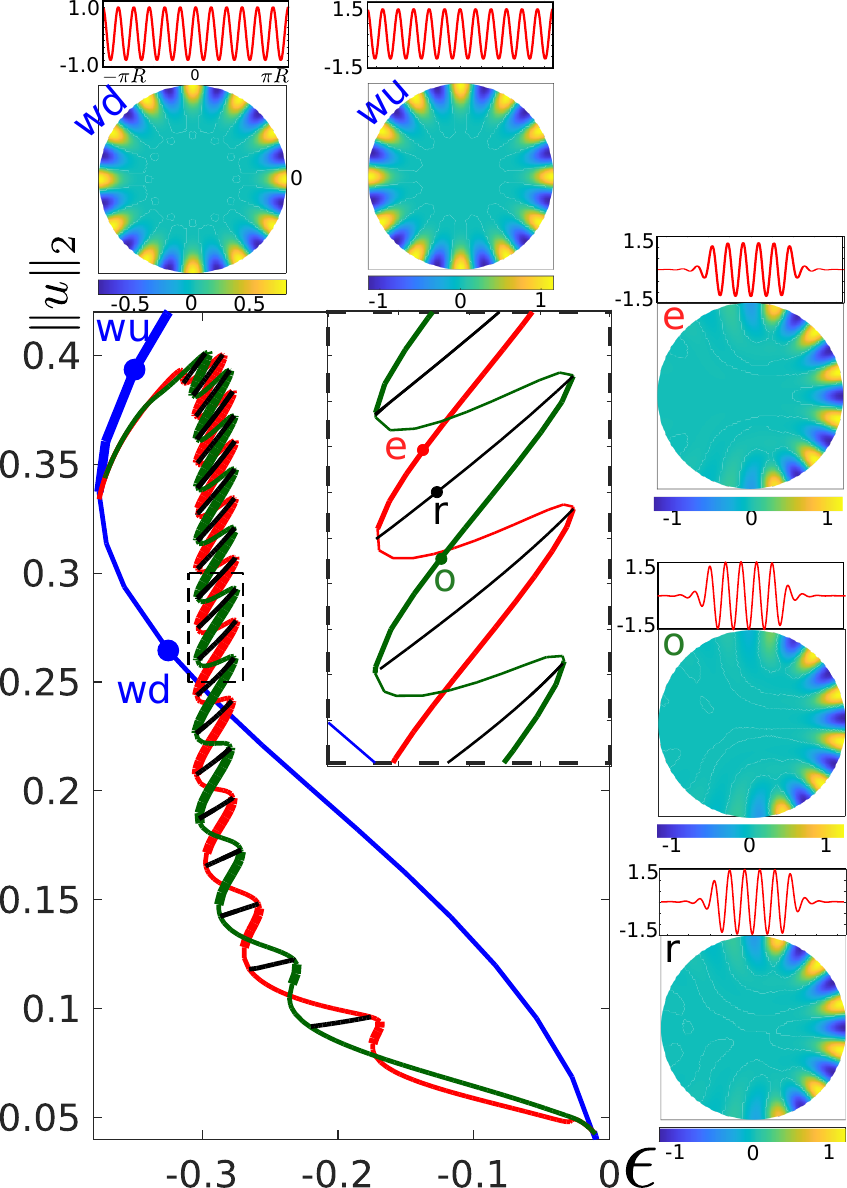}
    \end{center}
    \caption{The wall modes or daisy states (blue) and their secondary
      bifurcations computed from Eq.~(\ref{eq:sh35}) when $R=14$, $\nu=1.4$, and
      $q=1$. The secondary branches of (e)ven (red) and (o)dd (green) localized
      states are connected via (r)ung states (black). A zoom
      of the bifurcation diagram provides more detail of the localized
    branches. Representative solutions marked with dots in the
    bifurcation diagram are represented in the respective insets. 
    The 1D profiles correspond to each solution
    along its periphery.
    }
    \label{fig:daisy}
    \end{figure}
  

%% file: df2.tex
    \begin{figure}
    \begin{center}
    \ig[width=8.6cm]{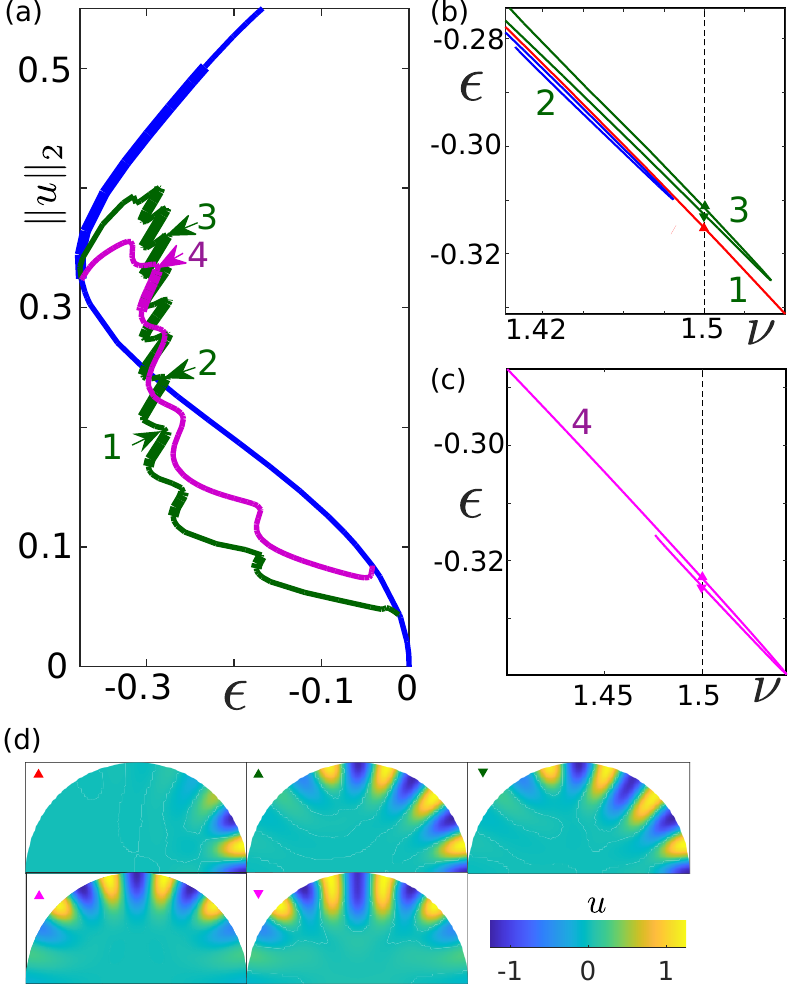}
  \end{center}
  \caption{Branches of one-
    and two-pulse solutions on the half-disk for the case
    $R=14,q=1$, together with continuation of selected folds in $\nu$. 
    (a) Bifurcation diagram of the daisy branch (blue), together with the
    branches of 1-pulse (green) and 2-pulse (magenta) localized daisy states
    when $\nu=1.4$, with four representative folds
    numbered 1--4.  
(b) Continuation of folds 1--3
    on the 1-pulse branch in the $(\nu,\epsilon)$ plane. 
    The numbers next to each curve correspond
    to the folds in (a); states at $\nu=1.5$ are marked by triangles. 
    (c) Same as (b) for the 2-pulse
    folds. (d) Five representative solutions from (b) and (c).}
    \label{fig:stripdaisy}
  \end{figure}


%% file: df3.tex
 \begin{figure}
    \begin{center}
    \ig[width=8.6cm]{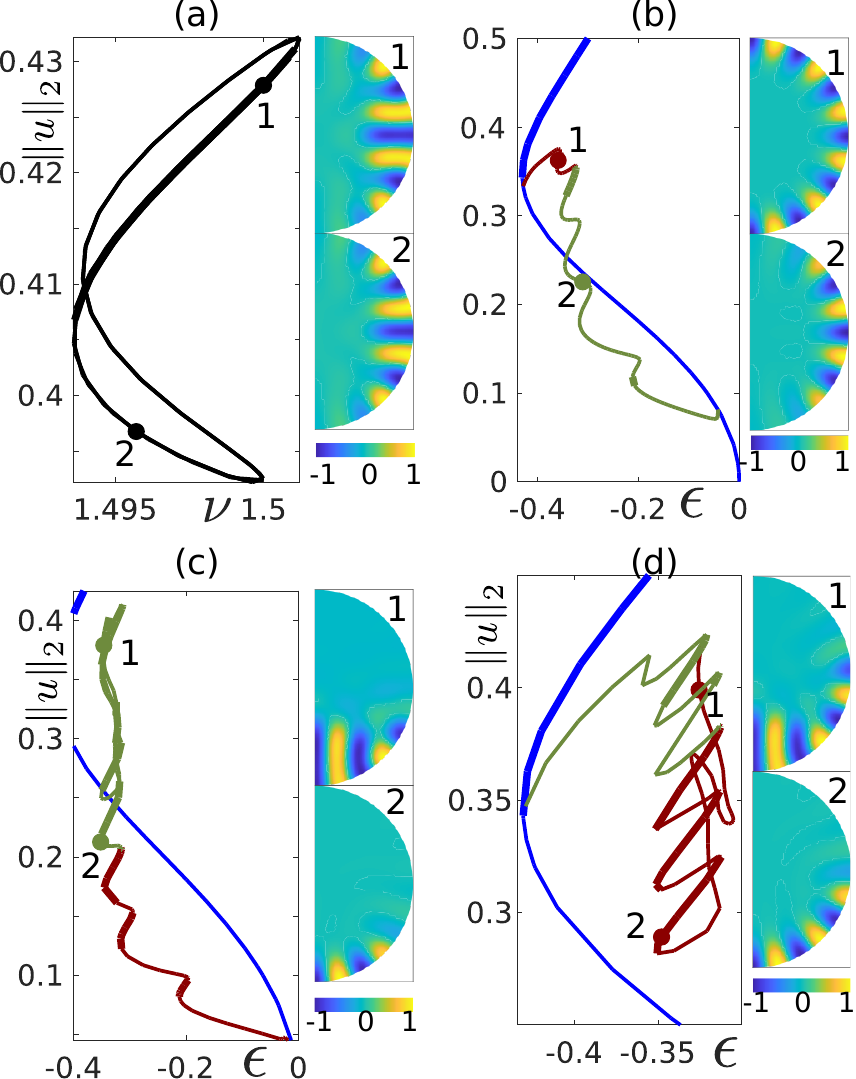}
  \end{center}
  \caption{Continuation of representative
    states in Fig.~\ref{fig:stripdaisy}(d)
    in the parameter (a) $\nu$, starting from the downwards
    pink triangle, (b) $\epsilon$, starting from the
    upwards pink triangle, (c) $\epsilon$, starting
    from the red upwards triangle, and (d) $\epsilon$, starting from
    the upwards green triangle.
    In (b)--(d), green (red) correspond to continuation for
    increasing (decreasing) $\epsilon$. Profiles with representative solutions
    are included.}
  \label{fig:nu15}
  \end{figure}

%% file: wf1.tex
\begin{figure}
  \begin{center}
\btab{l}{{\sm (a)}\\
  \hs{-3mm}\ig[width=40mm]{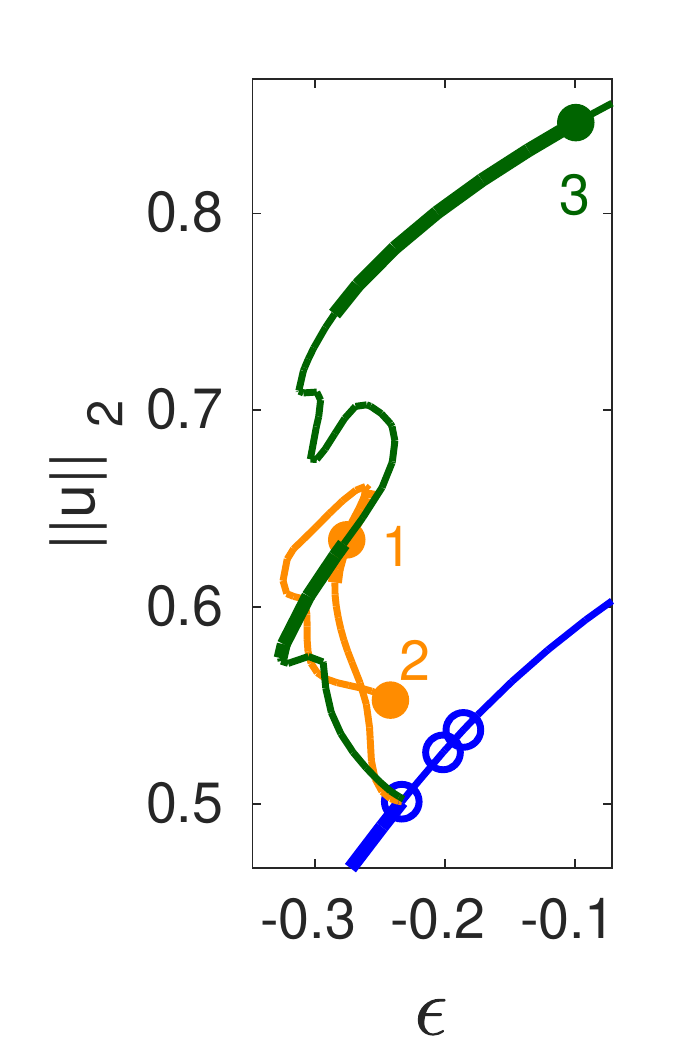}
\rb{27mm}{\btab{l}{\ig[width=22mm]{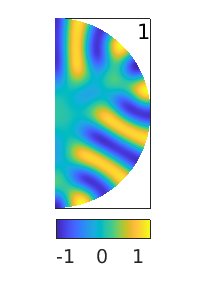}
\ig[width=22mm]{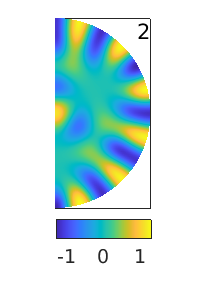}\\[-3mm]
\ig[width=22mm]{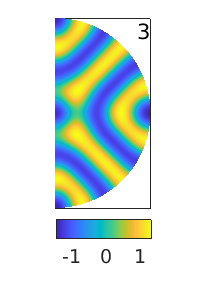}
\ig[width=22mm]{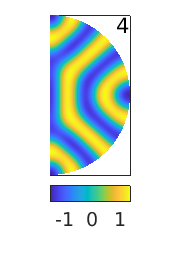}}}\\
{\sm (b)}\\
  \ig[width=35mm]{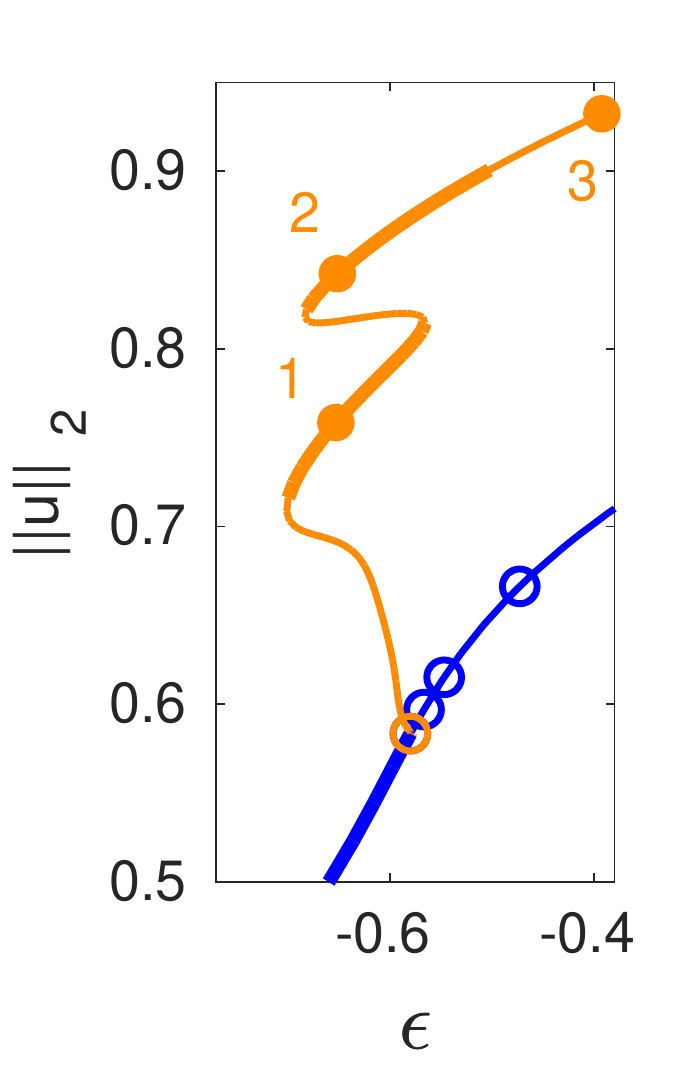}
\rb{25mm}{\btab{l}{\ig[width=22mm]{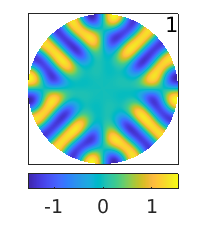}
\ig[width=22mm]{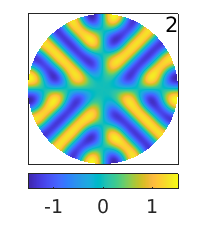}\\
\ig[width=22mm]{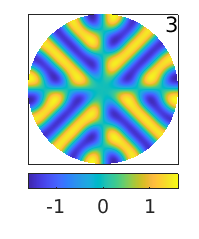}
\hs{-0mm}\rb{0mm}{\ig[width=25mm]{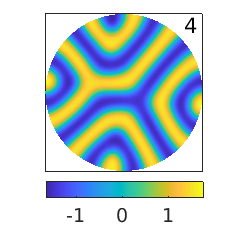}}}}\\
}

  \end{center}

\vs{-5mm}
\caption{Branches bifurcating from the daisy branch (blue) when it loses
stability at large amplitude. (a) $\nu=1.4$. The first bifurcating branch (orange) 
initially consists of
stable wall-attached 3-worms, which disintegrate upon further continuation 
into various combinations of boundary and center spots but never again become
stable. The next branch consists of wall-attached 4-worms (green) which 
turn into a combination of stripes perpendicular 
to the wall, spots in between, and a hole at the disk center,
with intervals of stability up to $\eps\approx -0.1$. State 4 is generated
from DNS starting from state 3, slightly to the right of the stability 
range of the green branch. 
(b) $\nu=2$. The first bifurcating branch is now a snaking branch of
wall-attached 4-worms similar to those in (a); state 4 is again obtained
from DNS starting from state 3. 
    \label{wf1}}
  \end{figure}

%% file: paf1.tex
\begin{figure}
  \begin{center}
  \hs{-3mm}\ig[width=38mm]{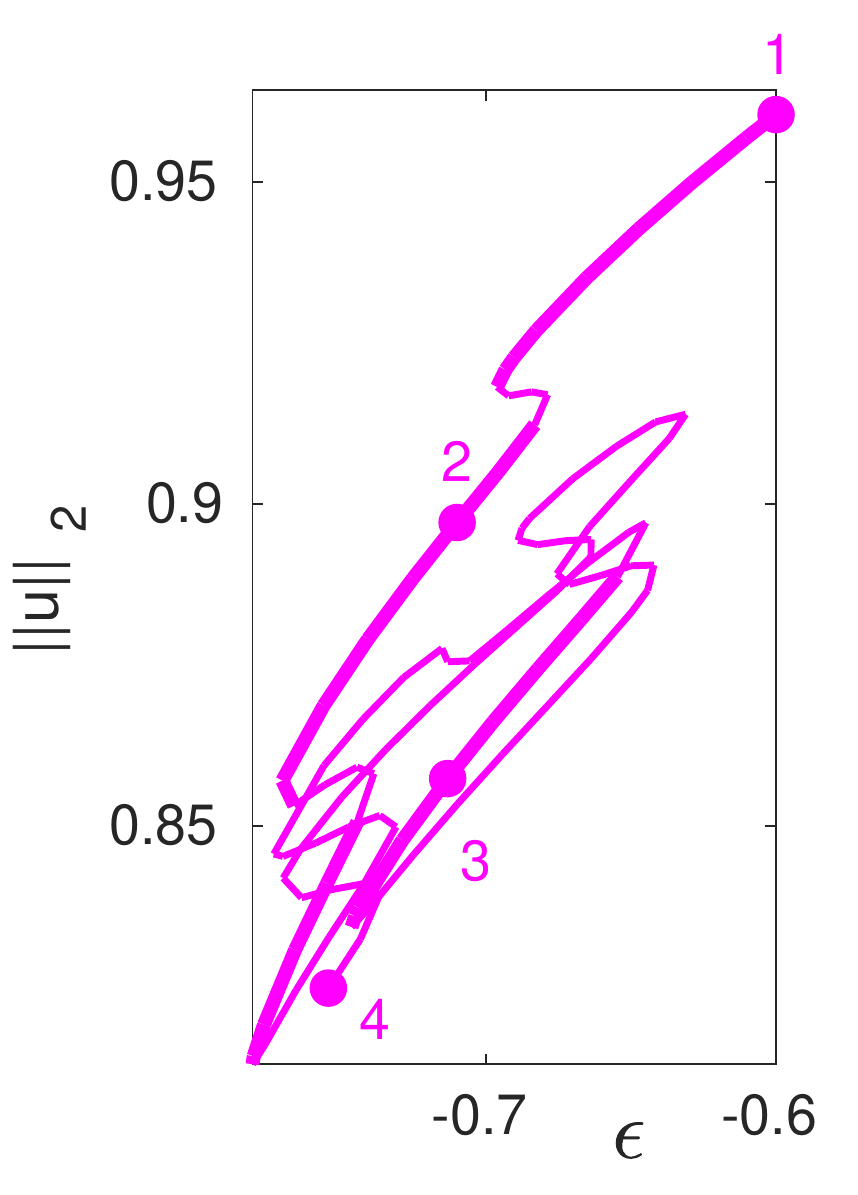}
\hs{-3mm}\rb{24mm}{\btab{l}{
\ig[width=27mm]{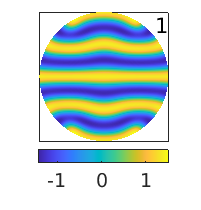}
\hs{-3mm}\ig[width=27mm]{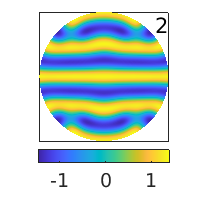}\\[-3mm]
\ig[width=27mm]{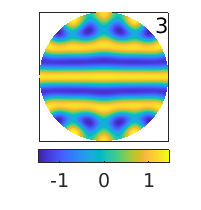}
\hs{-3mm}\ig[width=27mm]{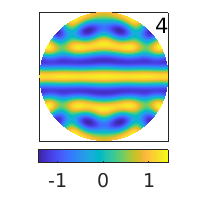}}}

  \end{center}

\vs{-5mm}
  \caption{Patterns for $\nu=2$ obtained from DNS with straight horizontal 
stripes as initial condition (state 1), and subsequent continuation 
in $\eps$.    \label{paf1}}
  \end{figure}